\documentclass[twocolumn,prl,aps,groupedaddress]{revtex4}
\usepackage{graphicx}
\usepackage{xcolor}
\usepackage{graphicx}
\usepackage{amsmath}
\usepackage{bm}
\usepackage{epsf,epsfig,graphics}
\usepackage{float}
\usepackage{makeidx}
\usepackage{latexsym}
\usepackage{amssymb}
\usepackage{mathrsfs}
\usepackage{amsbsy}
\usepackage{subfigure}

\usepackage{color}
\usepackage{morefloats}
\DeclareMathAlphabet\mathbfcal{OMS}{cmsy}{b}{n}
\usepackage{makecell,tabularx,booktabs}
\usepackage{textcomp}
\usepackage{xcolor}
\usepackage{lineno}
\makeatletter
\usepackage{textcomp}
\usepackage{hyperref}
\hypersetup{colorlinks=true, citecolor=purple, urlcolor=brown, linkcolor=blue}
\begin{document}

\title{Theory of Non-equilibrium Asymptotic State Thermodynamics: Interacting Ehrenfest Urn Ring as an Example}

\author{Chi-Ho Cheng  $^{1}$}
\email{phcch@cc.ncue.edu.tw}
\author{Pik-Yin Lai   $^{2,3}$}
\email{pylai@phy.ncu.edu.tw}

\affiliation{$^{1}$Department of Physics, National Changhua University
of Education, Taiwan, R.O.C.}
\affiliation{$^{2}$Department of Physics and Center for Complex Systems, National Central University, Taiwan, R.O.C.}
\affiliation{$^{3}$Physics Division, National Center for Theoretical Sciences, Taipei 10617, Taiwan, R.O.C.}
\date{\today}

\begin{abstract}
A generalized class of non-equilibrium state, called non-equilibrium asymptotic state (NEAS), is proposed.
The NEAS is constructed within the framework of the Fokker-Planck equations in thermodynamic limit.
Besides the usual equilibrium state and non-equilibrium steady state (NESS), the class of NEAS could also
cover non-equilibrium periodic state (NEPS) in which its dynamics shows periodicity, non-equilibrium quasi-periodic state (NEQPS),
and non-equilibrium chaotic state (NECS) in which its dynamics becomes chaotic.
Based on the theory of NEAS thermodynamics, the corresponding thermodynamics of
different NEAS could also be determined.
Finally the interacting Ehrenfest urn ring model is used as an example to illustrate
how different kinds of NEAS (equilibrium state, uniform NESS, non-uniform NESS, NEPS) in three-urn case are identified in our framework.
In particular, the thermodynamics of NEPS and its phase transitions to other types of NEAS are studied.
\end{abstract}

%\pacs{03.75.Fi,05.30.Jp,32.80.Pj}
\maketitle

%----------------------------------------------------------------------
\section{I. Introduction}

In nature, equilibrium physics are widely studied, but non-equilibrium phenomena often occur.
The appearance of non-equilibrium may not be due to imperfect environment or by accident.
In some cases, non-equilibrium phenomena appear for the sake of some functionality, especially
just after breaking the detailed balance. For example, in recent study, it was found that
detailed balance violations in brain increase with physical and cognitive exertion \cite{brain}.

Near equilibrium, linear response theory provides a way to calculate the non-equilibrium properties
from the correlation functions at equilibrium. Beyond the linear response regime, in small systems,
fruitful results were obtained from the realm of stochastic thermodynamics in which
the thermodynamic quantities like heat transfer, external work done, entropy production
are defined along individual trajectories and also their relationship can be identified
\cite{sekimoto,seifert1,seifert2,seifert3,murashita}.
 The fluctuation theorem was first discovered in non-equilibrium steady states \cite{evan}, and were later proved in other non-equilibrium  situations \cite{gallavotti,kurchan,lebowitz}.
By relating free energy difference between two equilibrium states through non-equilibrium process with work done,  the Jarzynski relation was achieved \cite{jarzynski1,jarzynski2}.
Later on, many variations and its realization were proposed \cite{crooks1,crooks2,hummer}.
If the  initial and final equilibrium states are extended to non-equilibrium steady states, the Hatano-Sasa equality  was found \cite{hatano} which follows from a more general framework of steady state thermodynamics \cite{oono,sasa}.

Among the non-equilibrium states, the non-equilibrium steady states are widely studied \cite{lax,derrida,qian,esposito}.
Besides that, the non-equilibrium non-steady states may also induce lots of physical insight to non-equilibrium physics.
For examples, when the system exhibits periodic dynamics, its thermodynamics may behave differently.
Recently, a model of three state interacting driven oscillators is shown to exhibit periodic dynamics.
After incorporating with stochastic thermodynamics process, phase transition between different non-equilibrium
states is illustrated \cite{herpich18}. Under the framework of stochastic thermodynamics, the driven Potts model
also exhibits the thermodynamics of non-equilibrium non-steady states \cite{herpich19}.
The inclusion of noise (similar to temperature effect)
into the standard Stuart-Landau dimer model \cite{ryu} (its dynamics exhibits limit cycles)
provides the insight into the importance of coherent synchronization within the working
substance in the operation of a thermal machine.

Even in the quantum case, limit cycles induced by
periodically driven quantum thermal machines may provide new insights towards the development of quantum
thermodynamics \cite{lazarides,brandner,diermann,schmidt,ikedal,liu,pascual}.

Throughout the above studies, there is a lack of formulation to merge dynamics and thermodynamics
in a natural way. Within this formulation, in thermodynamic limit,
macroscopic quantities and thermodynamic phenomena like phase transition can be well defined and studied.
Their dynamics and thermodynamics would not be independent but mutually affected.
Our approach based on the Fokker-Planck equation is particularly
advantageous to handle systems in the thermodynamic limit (large N limit)
over other approaches.
This is the scope of the current manuscript.

The content is organized as follows. In section II, we first develop a formalism (large number of degree of freedom) to
study the asymptotic behavior of the thermodynamic state, namely the non-equilibrium asymptotic state (NEAS).
Within the framework, NEAS is identified as a WKB solution which covers equilibrium state, non-equilibrium steady state (NESS),
non-equilibrium periodic state (NEPS) in which its dynamics shows periodicity (limit cycles),
and even non-equilibrium chaotic state (NECS) in which its dynamics becomes chaotic.
After then, in section III, the Ehrenfest urn ring model \cite{cheng20,cheng21} is introduced
as an example to illustrate the NEAS. In particular, in section IV, when the number of urns is restricted to three,
the behaviors of different kinds of NEAS (equilibrium state,  uniform NESS, non-uniform NESS, NEPS) and their phase transitions are demonstrated.
In section V, the thermodynamic fluctuation effect
is retrieved from the fluctuation around the WKB solution to the Fokker-Planck equation.
A thermodynamic relation is then identified.
To characterize the nature of phase transition,
a correspondence between dynamical and thermodynamic characterization is found.
Further it can also be proved that the dynamical stability criteria implies the thermodynamic stability.
Incidentally, both criteria are equivalent in NESS of the Ehrenfest ring model.
Finally, the conclusion is presented in section VI.

\section{II. Formalism of Non-equilibrium Asymptotic States}

In this section, we show how one extends the concept of non-equilibrium steady state (NESS) to
non-equilibrium asymptotic state (NEAS). The framework for NEAS also holds for NESS, i.e.,
NESS is just a special case in the NEAS formalism.

Suppose we consider the Fokker-Planck equation of dimension $D$ of the following form
\begin{widetext}
\begin{eqnarray} \label{fpe1}
   \frac{\partial \rho(\vec x,t)}{\partial t}
  = -\sum_{i=1}^D \frac{\partial}{\partial x_i}[A_i(\vec x,t) \rho(\vec x,t)]
    + \frac{1}{2 N}\sum_{i,j=1}^D \frac{\partial^2}{\partial x_i \partial x_j}
  [B_{ij}(\vec x,t) \rho(\vec x,t)]
\end{eqnarray}
\end{widetext}
where $\rho(\vec x, t)$ is the probability density of state $\vec x$ in our system.
$A_i(\vec x,t) \equiv \lim_{\tau\rightarrow 0}\frac{1}{\tau}\int d^D x' (x'_i-x_i) W(\vec x',t+\tau|\vec x,t)$
and
$B_{ij}(\vec x, t)\equiv N \lim_{\tau\rightarrow 0}\frac{1}{\tau}\int d^D x' (x'_i-x_i) (x'_j-x_j)W(\vec x',t+\tau|\vec x,t)$
with the transition probability $W(\vec x',t+\tau|\vec x,t)$ from state $\vec x$ at time $t$ to state $\vec x'$ at time $t+\tau$.
$N$ is a large number proportional to the system's degrees of freedom.
(For example, in a system of many particles, $N$ refers to the total particle number.)
$A_i(\vec x,t)$ and $B_{ij}(\vec x,t)$ are functions of $O(1)$.
Eq.(\ref{fpe1}) is a general equation describing the evolution of the probability density in large $N$ limit, which can be derived from the general master equation as outlined in Ref.\cite{cheng21}.
In large $N$ limit (thermodynamic limit),
the WKB ansatz could be applied \cite{risken}, which reads
\begin{eqnarray} \label{saddle1}
\rho (\vec x,t) \propto {\rm e}^{N f}
\end{eqnarray}
where $f$ is a function of $O(1)$. Traditional WKB method takes $f=f(\vec x)$, which
means to assume that the final state is a steady state. It is beyond the scope of
the method if the steady state is not favorable (to non-steady states) or even it doesn't exist.
Hence we extend $f=f(\vec x)$ to $f=f(\vec x,t)$ so that the steady and non-steady states
are both considered.
Further, assume there exists
a time-dependent optimal point \cite{saddlepoint} ${\vec \xi}(t)$  such that $\partial_i f({\vec \xi}(t), t)=0$
(its stability could be justified later). Hence one can expand $f(\vec x, t)$ around
${\vec x}={\vec \xi}(t)$, {\rm i.e.}, $f(\vec x,t) \simeq f(\vec \xi(t),t)+\frac{1}{2} \sum_{ij} \partial_{ij}f(\vec\xi(t),t)(x_i-\xi_i(t))(x_j-\xi_j(t))$ so that Eq.(\ref{saddle1}) can be
re-written as
\begin{eqnarray} \label{saddle2}
\rho(\vec x, t) \propto \exp [ N \sum_{i,j=1}^D c_{ij}(t) (x_i-\xi_i(t)) (x_j-\xi_j(t)) ]
\end{eqnarray}
up to the leading order in $N$, and $c_{ij}(t) \equiv \frac{1}{2} \partial_{ij}f(\vec\xi(t),t)$.
Obviously the matrix $\bf c$ is symmetric by its definition.

The form of Eq.(\ref{saddle2}) implies the main contribution of probability density from
the neighborhood of $\vec x=\vec \xi(t)$ in large $N$ limit. We can then further simplify Eq.(\ref{fpe1}) by expanding
$A_i(\vec x,t) \simeq A_i(\vec \xi(t),t) + \sum_j \partial_j A_i(\vec \xi(t),t)(x_j-\xi_j(t))
\equiv A_i(\vec \xi(t),t) + \sum_j a_{ij}(t)(x_j-\xi_j(t))$, and
$B_{ij}(\vec x,t)\simeq B_{ij}(\vec \xi(t),t)\equiv b_{ij}(t)$.
Under this approximation, Eq.(\ref{fpe1}) becomes
\begin{widetext}
\begin{eqnarray} \label{fpe2}
   \frac{\partial \rho(\vec x,t)}{\partial t}
  = -\sum_{i=1}^D \frac{\partial}{\partial x_i}\left[ \left( A_i(\vec\xi(t),t)+\sum_{j=1}^D a_{ij}(t)(x_j-\xi_j(t))\right)
  \rho(\vec x,t)\right]
    + \frac{1}{2 N}\sum_{i,j=1}^D b_{ij}(t)  \frac{\partial^2 }{\partial x_i \partial x_j}
\rho(\vec x,t)
\end{eqnarray}
\end{widetext}
Substitution of Eq.(\ref{saddle2}) into the above equation and keeping the leading order in $N$
(See Appendix A for details) gives
\begin{eqnarray}
\partial_t\vec \xi &=& {\vec A}(\vec \xi,t) \label{dynamics1} \\
\partial_t {\bf c}^{-1} &=&   {\bf a} {\bf c}^{-1} + {\bf c}^{-1} {\bf a}^{\rm t} - 2 {\bf b}  \label{fluctutation1}
\end{eqnarray}
Eq.(\ref{dynamics1}) is the dynamical equation to describe the state evolution.
The asymptotic behavior of the dynamical state (here we call it $\vec \xi^{\rm as}(t)$) could be
fixed points ($\partial_t \vec \xi^{\rm as} = 0$),
limit cycles ($\vec \xi^{\rm as}(t)$ follows some kind of periodic trajectory), quasi-periodic
or even chaotic states \cite{hirsch}.

After incorporating with thermal fluctuations from Eq.(\ref{fluctutation1}),
the thermodynamic states are then classified into equilibrium state (dynamical fixed point with
detailed balance), non-equilibrium steady state (NESS, dynamical fixed point with detailed balance violation),
non-equilibrium periodic state (NEPS, dynamical limit cycles), non-equilibrium quasi-periodic state (NEQPS) and non-equilibrium chaotic state (NECS).
All the above thermodynamic states are the possible non-equilibrium asymptotic state (NEAS), from which we define
the asymptotic behavior of Eqs.(\ref{dynamics1})-(\ref{fluctutation1}) in the framework under Eq.(1) with WKB ansatz in Eq.(\ref{saddle1}).

For NESS, there are two stability criteria. One is the dynamical stability derived from Eq.(\ref{dynamics1}),
saying that the real part of all eigenvalues of $\bf a$ is negative.
The other is the thermodynamic stability which is justified by its (thermodynamic) flucutation $\bf c$, in which its time
evolution is described in Eq.(\ref{fluctutation1}). In particular, at NESS which
corresponds to fixed point $\partial_t \vec \xi=0$, $\bf c$ is time independent. From
Eq.(\ref{fluctutation1}), $\partial_t {\bf c}=0$ leads to the Lyapunov equation
$ {\bf a} {\bf c}^{-1} + {\bf c}^{-1} {\bf a}^{\rm t} = 2 {\bf b}$ (Please refer to Appendix A
in Ref.\cite{cheng21} for details). If further at equilibrium, the detailed balance condition (${\bf a b} = {\bf b} {\bf a}^{\rm t}$)
 is satisfied,
$\bf c = b^{-1} a$
(Please refer to Appendix B in Ref.\cite{cheng21} for details).
The NESS stability condition is that all eigenvalues of $\bf c$ are negative.
It can be proved that, for NESS, the thermodynamic stability is satisfied if the system is dynamical stable
(See Appendix B for the proof).

\section{III. Ehrenfest Urn Ring Model}

\begin{figure}[h]
  \begin{center}
    \includegraphics[width=3in]{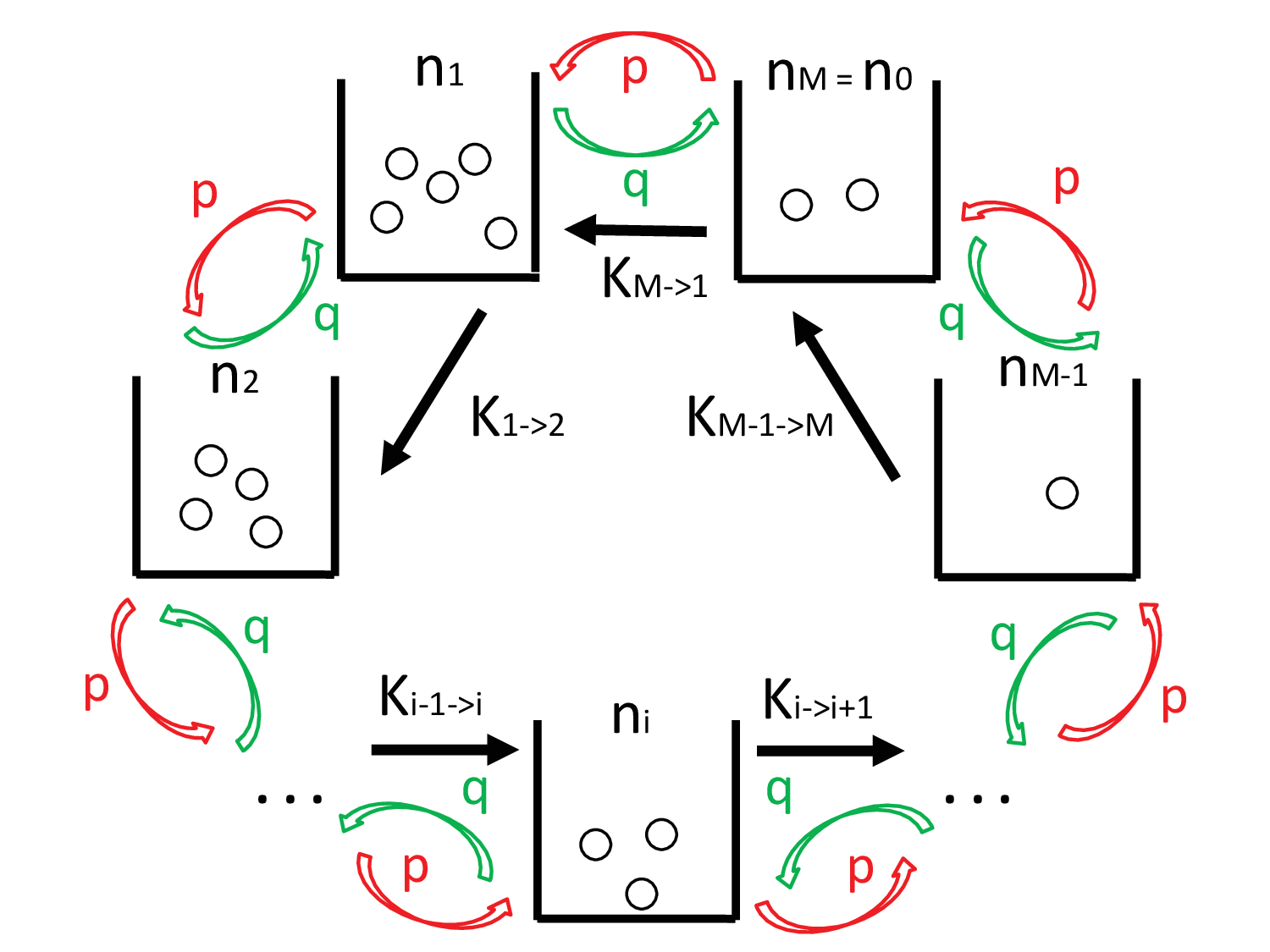}
  \end{center}
  \vspace{-5pt}
  \caption{Schematic diagram of the interacting Ehrenfest urn ring model. $M$ urns with particle numbers
    $n_i$ in the $i$-th urn are connected in a ring. For convenience, we label $n_0 \equiv n_M$.
    The direct jumping rate in anticlockwise (clockwise) direction is $p$ ($q$). $K_{i\rightarrow j}$ represents
    net particle flow rate from the $i$-th to the $j$-th urn.}
  \label{schematic.eps}
  \vspace{25pt}
\end{figure}

In this section, we applied the NEAS formalism in previous section on the Ehrenfest urn ring model with interactions \cite{cheng20,cheng21} to
illustrate the NEAS thermodynamics. As shown in Fig.\ref{schematic.eps}, $M$ urns are connected in a ring.
Particles in the same urn interact with each other, but no interaction between two particles
from different urns. Further, a direct jumping rate along the ring is introduced such that
the probability of anticlockwise (clockwise) direction is $p$ ($q$), and $p+q=1$ is imposed for
convenience.

The state of the system is labeled by the particle occupation distribution in the urn ring,
$\vec n=(n_1, n_2, \dots, n_M)$ where $n_i$ is the particle number in the $i$-th urn
with fixed total particle number $N$, {\rm i.e.}, $\sum_{i=1}^M n_i= N$.
For convenience, we also define $n_0 \equiv n_M$ (Periodic Boundary Condition).
At each step,
the transition probability from state $\vec n$ to state $\vec m$ is
\begin{eqnarray}  \label{tnm}
  T_{\vec m,\vec n} &=&
  \frac{1}{{\rm e}^{-\frac{g}{N}(n_i-n_j-1)}+1}
\end{eqnarray}
where $m_i = n_i-1$ and $m_j = n_j+1$.
$g\equiv N J \beta$ where $\beta$ is the inverse of effective temperature
(Please refer to Ref.\cite{cheng17} for its derivation).
After $s$ steps
from the initial state, the state probability $\rho(\vec n, s)$ satisfies the following
master equation,
\begin{eqnarray} \label{master1}
  \rho(\vec n,s+1) - \rho(\vec n, s)
  = \sum_{\vec m} \left( W_{\vec n,\vec m} \rho(\vec m,s) - W_{\vec m, \vec n} \rho(\vec n,s) \right)
\end{eqnarray}
where the discrete transition probability from state $\vec n$ to state $\vec m$ is
$W_{\vec m, \vec n} = \frac{n_i}{N} p T_{\vec m, \vec n}$
for anti-clockwise jump
and
$W_{\vec m, \vec n} = \frac{n_i}{N} q T_{\vec m, \vec n}$
for clockwise jump.

Let the (physical) time $t= \frac{\tau_1}{N} s $, where
$\tau_1$ is the time scale of each single step from $s$ to $s+1$,
and $\tau_1 \equiv 1$ in the following for convenience.
$\vec x\equiv \vec n/N$.
In the large $N$ limit, the discrete master equation in Eq.(\ref{master1})
can be transformed into the Fokker-Planck equation in Eq.(\ref{fpe1})
with $D=M-1$
after we further keep terms up to $O(1/N^2)$. The corresponding $A_i(\vec x)$
and $B_{ij}(\vec x)$ are
\begin{widetext}
\begin{eqnarray}
  && A_i(\vec x) =
  - \frac{p x_i}{{\rm e}^{-g(x_i-x_{i+1})}+1}
  + \frac{q x_{i+1}}{{\rm e}^{-g(x_{i+1}-x_i)}+1}
  + \frac{p x_{i-1}}{{\rm e}^{-g(x_{i-1}-x_i)}+1}
  - \frac{q x_i}{{\rm e}^{-g(x_i-x_{i-1})}+1}    \label{Ai}  \\
  && B_{ii}(\vec x) =
    \frac{p x_i}{{\rm e}^{-g(x_i-x_{i+1})+1}}
  + \frac{q x_{i+1}}{{\rm e}^{-g(x_{i+1}-x_i)}+1}
  + \frac{p x_{i-1}}{{\rm e}^{-g(x_{i-1}-x_i)}+1}
  + \frac{q x_i}{{\rm e}^{-g(x_i-x_{i-1})}+1}      \\
  && B_{i,i+1}(\vec x) = B_{i+1,i}(\vec x) =
  - \frac{p x_i}{{\rm e}^{-g(x_i-x_{i+1})}+1}
  - \frac{q x_{i+1}}{{\rm e}^{-g(x_{i+1}-x_i)}+1}
\end{eqnarray}
\end{widetext}
which do not have explicit time dependence, i.e.  the system is autonomous with ${\vec A}={\vec A}({\vec \xi}(t))$ in Eq.(\ref{dynamics1}).
Notice that the dimension of the state $\vec x$ in the probability density $\rho(\vec x,t)$
is reduced to $M-1$ because there is only $M-1$ independent variables due to the constraint $\sum_{i=1}^M x_i = 1$.
The population fraction of the particles are in general heterogeneous and time-dependent. One can define the non-uniformity for arbitrary $M$ urns \cite{cheng21}
\begin{eqnarray} \label{nonuniformity}
\psi(t) \equiv \frac{1}{M(M-1)} \sum_{(i<j)=1}^M \langle (x_i - x_j)^2 \rangle
\end{eqnarray}
which can also reflect to some extend the orderliness of the non-equilibrium state.

\subsection{Non-equilibrium Thermodynamic Relation}
We identified the thermodynamic relation relating the entropy production ($\frac{dS}{dt}$) and internal entropy production ($\frac{d_iS}{dt}$) rates to the work ($\frac{dW}{dt}$) and energy ($\frac{dE}{dt}$) rates of the system:
\begin{eqnarray} \label{thermolaw}
\frac{dS}{dt} = \frac{d_iS}{dt} + \beta \frac{dE}{dt} + \beta \frac{dW}{dt},
\end{eqnarray}
which is proved to be valid for the $M$-urn ring model in Appendix D for general non-equilibrium processes (asymptotic or non-asymptotic states).
The thermodynamic relation has been verified in the NESS for the 3-urn model in Ref.\cite{cheng21}, and the relevant energetic quantities of the NEPS in the $M$-urn  ring will be calculated below.

\subsection{Energetics of NESS and NEPS in the $M$-urn  ring}
When the system is at NEPS, suppose the oscillation period is $T$.
Consider the cyclic permutation symmetry of $M$ urns in a ring,
and then expand $\vec \xi^{\rm ps}(t)$ in Fourier series in time, which gives
\begin{eqnarray}  \label{xi}
\xi^{\rm ps}_i(t) = \frac{1}{M} + \sum_{k=1}^\infty c_k \cos( \frac{2 k \pi}{T}t - (i-1)\frac{2\pi}{M} )
\end{eqnarray}
where the coefficients $c_k$ depends on $g$ and $p$.

The (Boltzmann) entropy $S^{\rm ps}$ and the system energy $\beta E^{\rm ps}$ are given by
\begin{eqnarray}
S^{\rm ps} &\equiv& - \int d^{M-1}x  \rho^{\rm ps}(\vec x,t) \log\left( \rho^{\rm ps}(\vec x,t) / \frac{N!}{\prod_{i=1}^M n_i!} \right)
\nonumber \\
&=& -N \sum_{i=1}^M \xi^{\rm ps}_i(t) \log \xi^{\rm ps}_i(t) + O(1) \\
\beta E^{\rm ps} &\equiv&  \int d^{M-1}x \rho^{\rm ps}(\vec x,t) \frac{g}{2} \sum_{i=1}^M n_i(n_i-1)  \nonumber \\
&=& \frac{N g}{2} \sum_{i=1}^M (\xi^{\rm ps}_i(t))^2 + O(1)
\end{eqnarray}

Their rates of change are then
\begin{eqnarray} \label{dSdt}
\left.\frac{dS}{dt}\right|_{\rm ps} &=& -N \sum_{i=1}^M (\log\xi_i^{\rm ps}(t) + 1)A_i(\vec \xi^{\rm ps}(t)) \\
\label{dEdt}
\beta \left.\frac{dE}{dt}\right|_{\rm ps} &=& N g \sum_{i=1}^M \xi_i^{\rm ps}(t) A_i(\vec \xi^{\rm ps}(t))
\end{eqnarray}

Rate of work done by the system is
\begin{eqnarray}  \label{dWdt1}
\beta \left.\frac{dW}{dt}\right|_{\rm ps} &\equiv&
- N  \sum_{\vec n} \left\{ {\sum_{\vec m}}^{\rm ac}
  W_{\vec m, \vec n}\rho^{\rm ps}(\vec n,t) \log(\frac{p}{q}) \right. \nonumber \\
&& \left. + {\sum_{\vec m}}^{\rm c}
 W_{\vec n, \vec m}\rho^{\rm ps}(\vec m,t) \log(\frac{q}{p})  \right\}
\nonumber \\
&=& -\beta N \mu \sum_{\vec n} {\sum_{\vec m}}^{\rm ac}
( W_{\vec m, \vec n}\rho^{\rm ps}(\vec n,t) - W_{\vec n, \vec m}\rho^{\rm ps}(\vec m,t) ) \nonumber \\
&=& -\beta\mu \sum_{i=0}^{M-1} K_{i\rightarrow i+1}^{\rm ps}(t)
\end{eqnarray}
where ac (c) stands for anti-clockwise (clockwise) direction.
$\mu\equiv \beta^{-1} \log(\frac{p}{q})$ is the effective chemical potential difference
to actively drive the particle from the $i$-th to the $(i+1)$-th urn.
$K_{i\rightarrow i+1}^{\rm ps}(t)$ is the net particle flow rate from the $i$-th to the $(i+1)$-th urn
at NEPS, defined as
\begin{eqnarray} \label{netflow1} \label{Kij}
K_{i\rightarrow i+1}^{\rm ps}(t) &\equiv& N \sum_{\vec n}
( W_{(n_i-1,n_{i+1}+1),(n_i,n_{i+1})} \nonumber \\
&& - W_{(n_i+1,n_{i+1}-1),(n_i,n_{i+1})} ) \rho^{\rm ps}(\vec n,t)   \nonumber \\
&=& N \frac{p \xi_i^{\rm ps}(t) {\rm e}^{g\xi_i^{\rm ps}(t)} - q \xi_{i+1}^{\rm ps}(t) {\rm e}^{g\xi_{i+1}^{\rm ps}(t)}}
{{\rm e}^{g\xi_i^{\rm ps}(t)}+{\rm e}^{g\xi_{i+1}^{\rm ps}(t)}} +O(1) \nonumber \\
\end{eqnarray}

At NEPS, $\left.\frac{dS}{dt}\right|_{\rm ps}$ and $\left.\beta\frac{dE}{dt}\right|_{\rm ps}$ do not vanish, in general.
However, their time average $\langle \ldots \rangle_t$ over a period $T$ should be zero because both $S^{\rm ps}$ and $E^{\rm ps}$ are functionals of $\vec\xi^{\rm ps}(t)$
which is periodic.
\begin{eqnarray}
\left\langle \left.\frac{dS}{dt}\right|_{\rm ps} \right\rangle_t
&=& \frac{1}{T} \int_0^T dt \left.\frac{dS}{dt}\right|_{\rm ps}  \nonumber \\
&=& \frac{1}{T} ( S(\vec \xi^{\rm ps}(T)) - S(\vec \xi^{\rm ps}(0)) ) \nonumber \\
&=& 0
\end{eqnarray}
Same argument also applies to $E^{\rm ps}$ so that $\langle \left.\frac{dE}{dt}\right|_{\rm ps} \rangle_t =0$.
According to the thermodynamic law (See Appendix D),
$dS = d_iS + \beta dE + \beta dW$,
we then arrive at a generalized relationship at NEAS,
\begin{eqnarray}
\left\langle \left.\frac{d_i S}{dt}\right|_{\rm as} \right\rangle_t
= -\beta\left\langle \left.\frac{dW}{dt}\right|_{\rm as} \right\rangle_t
\end{eqnarray}
From Eq.(\ref{dWdt1}),
\begin{eqnarray} \label{dWK}
\beta \left\langle\left.\frac{dW}{dt}\right|_{\rm ps}\right\rangle_t
= -\beta\mu \sum_{i=0}^{M-1} \langle K_{i\rightarrow i+1}^{\rm ps}(t) \rangle_t
\end{eqnarray}

From the dynamical equation in Eq.(\ref{dynamics1}), taking time average,
\begin{eqnarray}
\langle A_i(\vec \xi^{\rm ps}(t)) \rangle_t
= \frac{1}{T} \int_0^T dt \partial_t \xi^{\rm ps}_i(t) = 0
\end{eqnarray}
Notice from Eqs.(\ref{Ai}) and (\ref{netflow1}) that
$N A_i(\vec \xi^{\rm ps}(t))
=-K^{\rm ps}_{i\rightarrow i+1}(t) + K^{\rm ps}_{i-1 \rightarrow i}(t)$, which implies
$\langle K_{i\rightarrow i+1}^{\rm ps}(t) \rangle_t$ are equal to each other for any $i$.

At NEPS, let $K^{\rm ps} \equiv \langle K_{i\rightarrow i+1}^{\rm ps}(t) \rangle_t$, then
\begin{eqnarray}  \label{diSKneps}
\left\langle \left.\frac{d_i S}{dt}\right|_{\rm ps} \right\rangle_t
= - \beta \left\langle\left.\frac{dW}{dt}\right|_{\rm ps}\right\rangle_t
= \beta\mu M K^{\rm ps}
\end{eqnarray}
which is a more generalized relationship at NEPS to connect the time average of
internal entropy production rate, the rate of work done to the system, and the net particle flow.
Notice that the relationship also holds for NESS in which these three variables are already
time independent,
\begin{eqnarray}  \label{diSKness}
 \left.\frac{d_i S}{dt}\right|_{\rm ss}
= - \beta  \left.\frac{dW}{dt}\right|_{\rm ss}
= \beta\mu M K^{\rm ss}.
\end{eqnarray}

\section{IV. Three urn ring model ($M=3$)}
In this section, we discuss the three urn case because $M=3$ is the minimal urn number to illustrate the NEPS, one of the NEAS never discussed before (only equilibrium states and NESS in previous studies \cite{cheng20,cheng21}). And since the phase space is two-dimensional, NEQPS or NECS are not possible in this case.
\subsection{Phase diagram, NEPS, uniform and non-uniform NESSs in the 3-urn model}
\begin{figure}[h]
  \begin{center}
   \subfigure[]{  \includegraphics[width=\columnwidth]{phasediagnew2.eps}}
      \subfigure[]{  \includegraphics[width=.48\columnwidth]{limitcycles.eps}}
      \subfigure[]{  \includegraphics[width=.48\columnwidth]{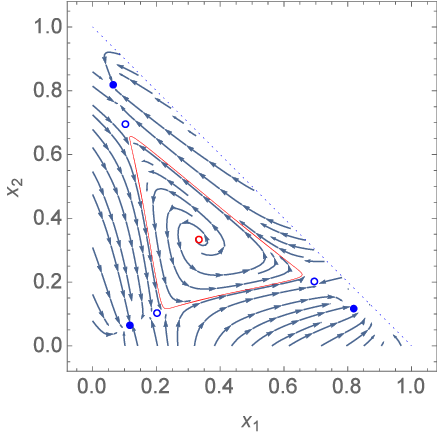}}
  \end{center}
  \vspace{-5pt}
  \caption{(a) Phase diagram of three urn ring model.
  There are five regions, uniform non-equilibrium steady state (uNESS),
  non-uniform non-equilibrium steady state (nuNESS), co-existence of both uNESS and nuNESS (Coexist I),
  non-equilibrium periodic state (NEPS), and co-existence of both NEPS and nuNESS (Coexist II).
  Three routes to NEPS, path 1 (decreasing $g$ at $p=0.9$ from uNESS),
  path 2 (increasing $g$ at $p=0.9$ from nuNESS),
  and path 3 (increasing $p$ at $g=-3.3$ from nuNESS).
  (b)  The long-time trajectories in the NEPS for $p=0.8$ and $g=-3.2$, $3.01$. Very close to the Hopf bifucation point, the oscillation just emerges with very small amplitude ($g=-3.001$) is also shown.
  (c) The phase portrait in the Coexist II regime with $p=0.8$ and $g=-3.35$. The periodic trajectory (red closed curve) coexists with the stable fixed points of the nuNESSs (filled blue circles), the corresponding unstable fixed points are also shown by the open blue circles. The unstable uNESS fixed point is marked by an open red circle.}
  \label{phasediagnew.eps}
\end{figure}

The phase diagram of three urn case is obtained from solving Eq.(\ref{dynamics1}), which is shown in Fig.\ref{phasediagnew.eps}(a).
At large $g$, the system is at uniform NESS (uNESS). When the particle interaction becomes
more and more attractive, until $g<-3$, the uNESS becomes unstable.
(In general, for arbitrary $M$, the uNESS is unstable when $g<-M$. See Appendix C for details.)
The system is at a non-uniform NESS (nuNESS)
at small unbalanced jumping rate, $p \gtrsim \frac{1}{2}$. By increasing $p$ up to a critical value,
the nuNESS becomes dynamically unstable via saddle-node bifurcation, {\rm i.e.}, Eq.(\ref{dynamics1}) has no stable fixed point and the phase boundary (blue solid curve) has been determined analytically \cite{cheng21}.

For $-3.8 \lesssim g<-3$ and large enough $p$, the NESS (uNESS or nuNESS) is unstable to
the non-equilibrium periodic state (NEPS).
It can be identified by observing $\xi_i(t)$ and $K_{i\rightarrow i+1}(t)$ in time asymptotically. Furthermore, by carefully examining the NEPS, it is found that there is a coexistence regime (Coexist II region bounded by the solid blue curve and red dot-dashed curve in Fig.\ref{phasediagnew.eps}(a))  in which nuNESS coexists with NEPS. The phase boundary (red dot-dashed curve) at which the NEPS vanishes via an infinite-period global bifurcation is obtained from the numerical solution of Eq.(\ref{dynamics1}).

\begin{figure}[H]
  \begin{center}
   \subfigure[]{  \includegraphics[width=\columnwidth]{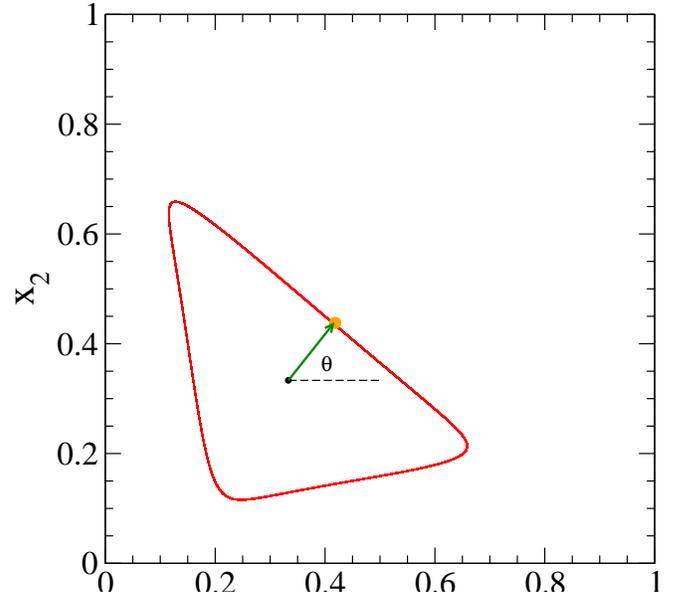}}
        \subfigure[]{  \includegraphics[width=\columnwidth]{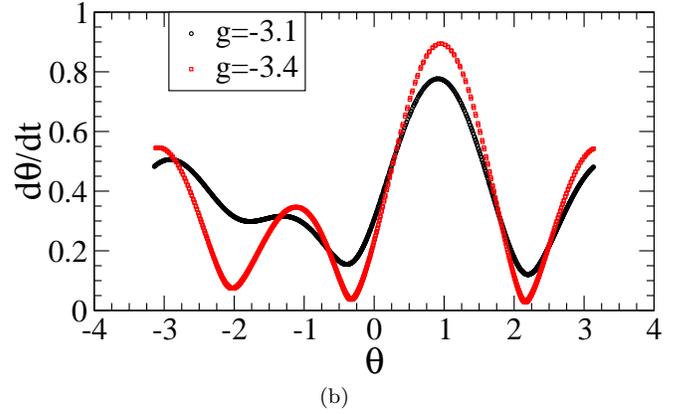}}
  \end{center}
\caption{(a) Periodic trajectory for the NEPS in the phase space with $p=0.9$ and $g=-3.4$ undergoing a counter-clockwise circulation. The instantaneous of $\vec{\xi}(t)$ is marked by the filled orange dot whose location relative to $(\frac{1}{3},\frac{1}{3})$ and can be represented by the angular variable $\theta(t)$ as shown. (b) $\dot{\theta}$ vs. $\theta$ in the NEPS for $p=0.9$ and $g=-3.1$ and $-3.4$.  }
  \label{thetadot}
\end{figure}

 The non-equilibrium phase transition between the uNESS and NEPS (the dashed vertical line phase boundary) such as path 1 in Fig.\ref{phasediagnew.eps}(a) is characterised by a Hopf bifurcation in the dynamical system from Eq.(\ref{dynamics1}).
 As $g$ decreases along path 1, the uNESS becomes unstable and gives ways to stable periodic dynamics at $g=-3^-$ with finite period and infinitesimally small emerging oscillation amplitude as shown in the phase space trajectories in Fig.\ref{phasediagnew.eps}(b),
 characterized by a supercritical Hopf bifurcation.  On the other hand, the non-equilibrium phase transition between the nuNESS and NEPS
 (the dot-dashed phase boundary in Fig.\ref{phasediagnew.eps}(a)) is characterised by an infinite-period global bifurcation.
 As demonstrated by the reverse path 3 in  Fig.\ref{phasediagnew.eps}(a), the system is in a Coexist II region in which NEPS and NESS coexist before arriving to the phase boundary. As shown in the phase portrait Fig.\ref{phasediagnew.eps}(c), the periodic NEPS trajectory (red closed curve) coexists with three pairs of stable (filled blue circles) and unstable (open blue circles) fixed points of the nuNESSs.
 Further decrease in $g$ will shift the periodic trajectory to be closer to the attracting manifold of the stable fixed point
 and hence increasing the oscillation period. Eventually when $g$ hits the phase boundary (red dot-dashed curve in Fig.\ref{phasediagnew.eps}(a)) and the trajectory falls onto the attractive fixed point and hence destroying the limit cycle,
 resulting in an infinite period global bifurcation. One can focus on the phase (angular) dynamics of the periodic dynamics of NEPS which can be characterized by the angular variable $\theta(t)$ as depicted in Fig.\ref{thetadot}(a). The effective phase dynamics can be obtained by  plotting  $\dot{\theta}$ vs. $\theta$ as shown in Fig.\ref{thetadot}(b).  As the system approaches the phase boundary (dot-dashed curve in Fig.\ref{phasediagnew.eps}(a)), the minima of $\dot{\theta}$ approach to zero, signifying the infinite period global bifurcation for the vanishing of the NEPS.  The period of the NEPS has a one-over-square-root divergence  following the classic scenario of critical slowing down in infinite period bifurcation \cite{strogatzbook}.

\subsection{Non-equilibrium fluxes and periodic oscillations}
\begin{figure}[H]
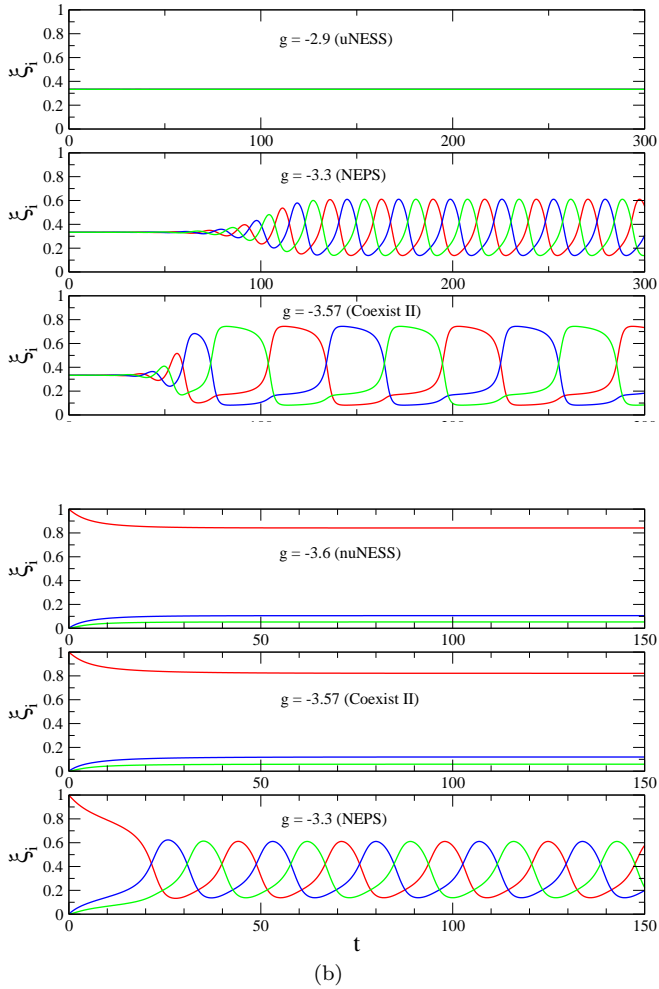

  \begin{center}
   \subfigure[]{  \includegraphics[width=\columnwidth]{urn3as01_p0d9_t_vs_xi_a.eps}}
   \subfigure[]{  \includegraphics[width=\columnwidth]{urn3as01_p0d9_t_vs_xi_b.eps}}
  \end{center}
  \caption{$\xi_i(t)$ are plotted as a function of $t$ at different $g$  with $p$=0.9.
  $\xi_1$, $\xi_2$, and $\xi_3=1-\xi_1-\xi_2$ are represented by red, blue, and green lines, respectively.
    The result is numerically solved from Eq.(\ref{dynamics1}).
  (a) The graphs correspond to the path 1 in Fig.\ref{phasediagnew.eps}(a)
  with initial condition $\vec \xi(0)=(\frac{1}{3},\frac{1}{3},\frac{1}{3})$.
  (b) The graphs correspond to the path 2 in Fig.\ref{phasediagnew.eps}(a)
  with initial condition $\vec \xi(0)=(1,0,0)$.
 }
  \label{urn3as01_p0d9_t_vs_xi.eps}
\end{figure}

\begin{figure}[H]
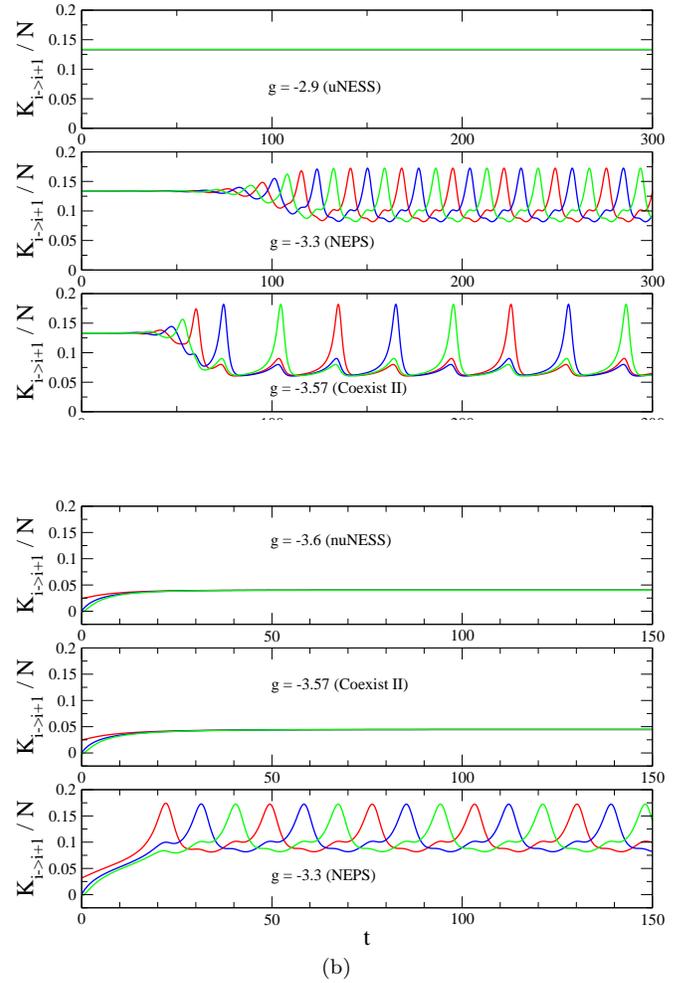

  \begin{center}
   \subfigure[]{  \includegraphics[width=\columnwidth]{urn3as01_p0d9_t_vs_Kij_a.eps}}
   \subfigure[]{  \includegraphics[width=\columnwidth]{urn3as01_p0d9_t_vs_Kij_b.eps}}
  \end{center}
  \caption{$K_{i\rightarrow i+1}(t)/N$ are plotted as a function of $t$ at different $g$  with $p$=0.9.
  $K_{1\rightarrow 2}$, $K_{2\rightarrow 3}$, and $K_{3\rightarrow 1}$(=$K_{0\rightarrow 1}$) are represented by red, blue, and green lines, respectively.
  The result is numerically solved from Eq.(\ref{netflow1}).
  (a) The graphs correspond to the path 1 in Fig.\ref{phasediagnew.eps}(a)
  with initial condition $\vec \xi(0)=(\frac{1}{3},\frac{1}{3},\frac{1}{3})$.
  (b) The graphs correspond to the path 2 in Fig.\ref{phasediagnew.eps}(a)
  with initial condition $\vec \xi(0)=(1,0,0)$.
  }
  \label{urn3as01_p0d9_t_vs_Kij.eps}
\end{figure}

To further examine the NEPS, three routes from NESS to NEPS, which are path 1, 2, 3 shown
in Fig.\ref{phasediagnew.eps}(a), are studied.
Figs.\ref{urn3as01_p0d9_t_vs_xi.eps} and \ref{urn3as01_p0d9_t_vs_Kij.eps} show
$\xi_i(t)$ and $K_{i\rightarrow i+1}(t)$ at different $g$ with fixed $p=0.9$, respectively.
The right and left column correspond to path 1 and 2, respectively.
It is seen that $\xi_i(t)$ and $K_{i\rightarrow i+1}(t)$ become periodic asymptotically at NEPS. In addition, the coexistence of nuNESS and NEPS (Coexist II regime) is explicitly demonstrated that for $g=-3.57$, nuNESS results in path 2 but path 1 leads to NEPS due to different different initial conditions.

%\begin{figure}[H]
%  \begin{center}
%    \includegraphics[width=\columnwidth]{urn3as01_p0d9_t_vs_xi.eps}
%  \end{center}
%  \vspace{-5pt}
%  \caption{$\xi_i(t)$ are plotted as a function of $t$ at different $g$  with $p$=0.9.
%  $\xi_1$, $\xi_2$, and $\xi_3=1-\xi_1-\xi_2$ are represented by red, blue, and green lines, respectively.
%  The initial conditions are $\vec \xi(0)=(1,0,0)$ in the left column and
%  $\vec \xi(0)=(\frac{1}{3},\frac{1}{3},\frac{1}{3})$ in the right column.
%  The result is numerically solved from Eq.(\ref{dynamics1}).
%  The graphs in the left (right) column correspond to the path 2 (1) in Fig.\ref{phasediagnew.eps}(a). }
%  \label{urn3as01_p0d9_t_vs_xi.eps}
%\end{figure}

%\begin{figure}[H]
%  \begin{center}
%    \includegraphics[width=\columnwidth]{urn3as01_p0d9_t_vs_Kij.eps}
%  \end{center}
%  \vspace{-5pt}
%  \caption{$K_{i\rightarrow i+1}(t)/N$ are plotted as a function of $t$ at different $g$  with $p$=0.9.
%  $K_{1\rightarrow 2}$, $K_{2\rightarrow 3}$, and $K_{3\rightarrow 1}$(=$K_{0\rightarrow 1}$) are represented by red, blue,
% and green lines, respectively.
%  The initial condition $\vec \xi(0)=(1,0,0)$ in the left column and
%  $\vec \xi(0)=(\frac{1}{3},\frac{1}{3},\frac{1}{3})$ in the right column.
%  The result is numerically solved from Eq.(\ref{netflow1}).
%  The graphs in the left (right) column correspond to the path 2 (1) in Fig.\ref{phasediagnew.eps}(a).  }
%  \label{urn3as01_p0d9_t_vs_Kij.eps}
%\end{figure}

\begin{figure}[H]
  \begin{center}
    \includegraphics[width=\columnwidth]{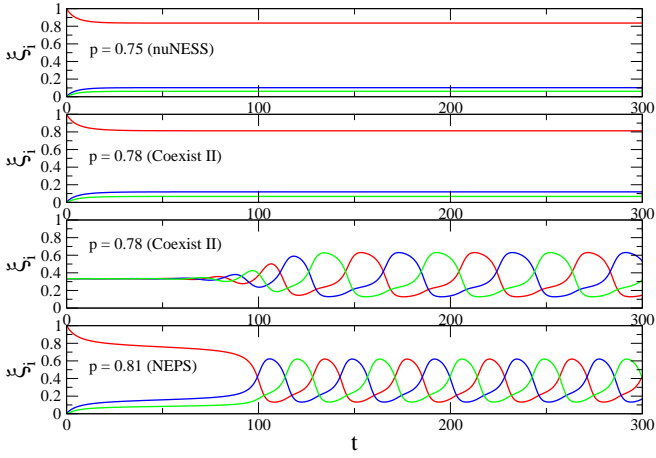}
  \end{center}
  \vspace{-5pt}
  \caption{$\xi_i(t)$ are plotted as a function of $t$ at $p=0.75, 0.78, 0.81$ with $g=-3.3$.
  $\xi_1$, $\xi_2$, and $\xi_3=1-\xi_1-\xi_2$ are represented by red, blue, and green lines, respectively.
  The initial conditions are $\vec \xi(0)=(1,0,0)$, except that it is $\vec \xi(0)=(\frac{1}{3},\frac{1}{3},\frac{1}{3})$ in the third row.
  The result is numerically solved from Eq.(\ref{dynamics1}).
  The graphs correspond to the path 3 in Fig.\ref{phasediagnew.eps}(a). }
  \label{urn3as01_gn3d3_t_vs_xi.eps}
\end{figure}

\begin{figure}[H]
  \begin{center}
    \includegraphics[width=\columnwidth]{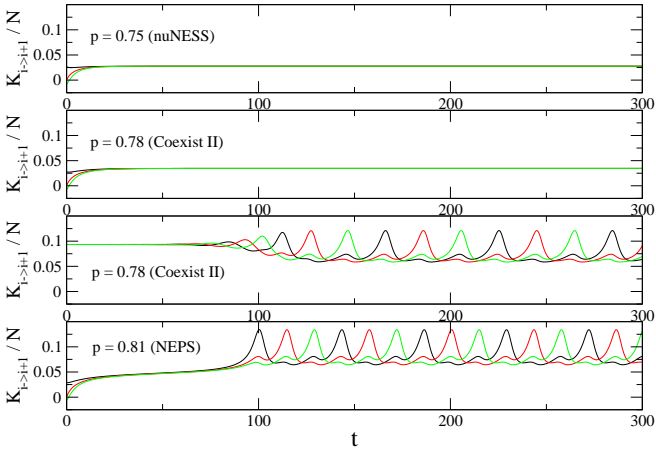}
  \end{center}
  \vspace{-5pt}
  \caption{$K_{i\rightarrow i+1}(t)/N$ are plotted as a function of $t$ at $p=0.75, 0.78, 0.81$  with $g=-3.3$.
  $K_{1\rightarrow 2}$, $K_{2\rightarrow 3}$, and $K_{3\rightarrow 1}$(=$K_{0\rightarrow 1}$) are represented by red, blue, and green lines, respectively.
  The initial conditions are $\vec \xi(0)=(1,0,0)$, except that it is $\vec \xi(0)=(\frac{1}{3},\frac{1}{3},\frac{1}{3})$ in the third row.
  The result is numerically solved from Eq.(\ref{netflow1}).
  The graphs correspond to the path 3 in Fig.\ref{phasediagnew.eps}(a).  }
  \label{urn3as01_gn3d3_t_vs_Kij.eps}
  \vspace{25pt}
\end{figure}

Figs.\ref{urn3as01_gn3d3_t_vs_xi.eps} and \ref{urn3as01_gn3d3_t_vs_Kij.eps} show
$\xi_i(t)$ and $K_{i\rightarrow i+1}(t)$ at different $p$ with fixed $g=-3.3$, respectively.
It corresponds to path 3. $\xi_i(t)$ and $K_{i\rightarrow i+1}(t)$ become periodic asymptotically arriving at the NEPS for large $p$.
Between nuNESS and NEPS, there is a coexistence region labeled by Coexist II.
The asymptotic behavior of $\xi_i(t)$ and $K_{i\rightarrow i+1}(t)$ will depend on the initial condition.
It can be a nuNESS or  NEPS asymptotically, as illustrated in the $p=0.78$ panels in Figs.\ref{urn3as01_gn3d3_t_vs_xi.eps} and \ref{urn3as01_gn3d3_t_vs_Kij.eps}.
%%%%%%%%%%%%%%%%%%%%%%%%%%%%%%%%%%%%%%%%%%%%%%%%%%%%%%%%%%%%%%%%%%%%%%%%%%%%%%%%%%%%%%%%%%%%%%%%%%%

\begin{figure}[H]
  \begin{center}
    \includegraphics[width=\columnwidth]{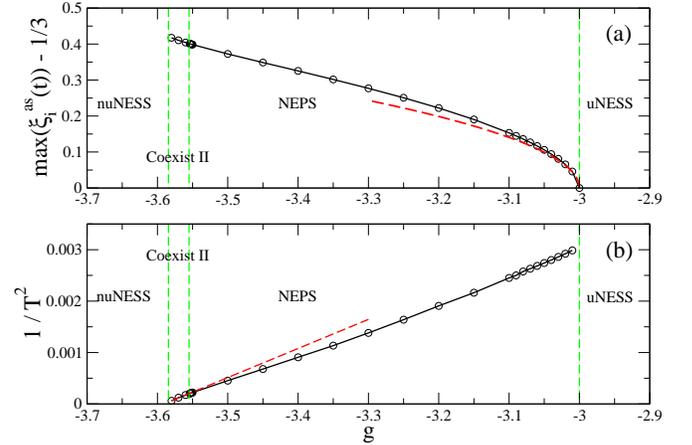}
  \end{center}
  \vspace{-5pt}
  \caption{(a) The amplitude of occupation oscillation, ${\rm max}(\xi^{\rm ps}_i(t))-\frac{1}{3}$,
  is plotted as a function of $g$ at $p=0.9$ at NEPS.
  Red dashed line is the analytical result near phase boundary from Eq.(\ref{etaas}).
  (b) The inverse of square of oscillation period, $1/T^2$, is plotted as a function of $g$ at $p=0.9$ at NEPS.
      Red dashed line is the linear fitting near phase boundary.
   The vertical dashed lines are the phase boundaries.}
  \label{urn3as01_p0d9_g_vs_amp_invT2.eps}
\end{figure}

\begin{figure}[H]
  \begin{center}
    \includegraphics[width=\columnwidth]{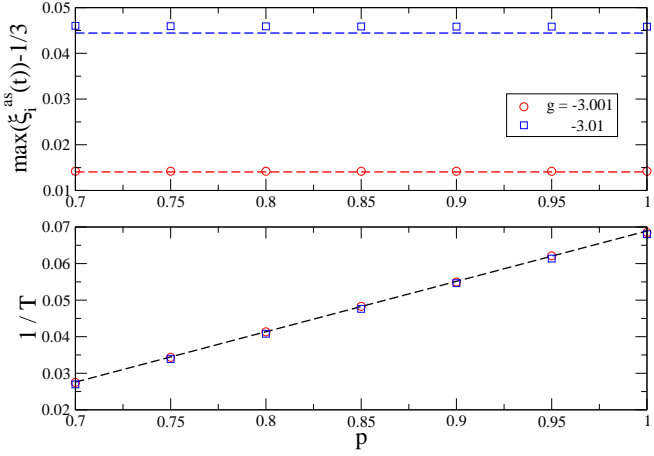}
  \end{center}
  \vspace{-5pt}
  \caption{(a) The amplitude of occupation oscillation, ${\rm max}(\xi^{\rm ps}_i(t))-\frac{1}{3}$,
  is plotted as a function of $p$ at $g=-3.001, -3.01$ at NEPS.
  Dashed lines are the corresponding analytical result from Eq.(\ref{etaas}).
  (b) The inverse of oscillation period, $1/T$, is plotted as a function of $p$ at $g=-3.001, -3.01$ at NEPS.
  The dashed line is the analytical result from Eq.(\ref{period}). }
  \label{urn3as01_u_p_vs_amp_invT.eps}
\end{figure}

As shown in Fig.\ref{urn3as01_p0d9_g_vs_amp_invT2.eps},
when the system from NEPS approaches the phase boundary of uNESS by increasing $g$ at fixed $p=0.9$ (reversed path 1),
the amplitude of occupation oscillation, ${\rm max}(\xi^{\rm ps}_i(t))-\frac{1}{3}$, drops to zero continuously whereas the
 corresponding oscillation period ($T$) decreases to a finite value. Such a behavior indicates clearly that the transition between uNESS and NEPS at $g=-3$ is via a supercritical Hopf bifurcation. Furthermore, for $g\rightarrow -3^-$, $|\xi^{\rm ps}_i(t) - \frac{1}{3}| \ll 1$, Eq.(\ref{dynamics1}) can be linearized as
\begin{eqnarray}
\frac{d}{dt} \left(
               \begin{array}{c}
                 \xi^{\rm ps}_1(t)-\frac{1}{3} \\
                 \xi^{\rm ps}_2(t)-\frac{1}{3} \\
               \end{array}
             \right)
= -\frac{1}{4}(p-q)
\left(
  \begin{array}{cc}
  1 & 2 \\
  -2 & -1 \\
  \end{array}
  \right)
\left(
\begin{array}{c}
\xi^{\rm ps}_1(t)-\frac{1}{3} \\
\xi^{\rm ps}_2(t)-\frac{1}{3} \\
\end{array}
\right)  \nonumber \\
\end{eqnarray}
which gives
\begin{eqnarray}
\frac{2\pi}{T} = \frac{\sqrt{3}}{4}(p-q)\quad \hbox{or } T = \frac{8\pi}{\sqrt{3} (p-q)} \label{period}
\end{eqnarray}
The above formula is in agreement with the numerical value
near phase boundary with $g \lesssim -3$, as shown in Fig.\ref{urn3as01_u_p_vs_amp_invT.eps}(b).

To capture the behavior of the amplitude of occupation oscillation, define
$\eta^{\rm ps}_i(t)\equiv \xi^{\rm ps}_i(t) - \frac{1}{3}$. Consider only the first harmonic component
in Eq.(\ref{xi}), one get
\begin{eqnarray}  \label{eta1}
\eta^{\rm ps}_1(t) &=& \eta^{\rm ps} \cos(\frac{2\pi}{T}t) \\
\eta^{\rm ps}_2(t) &=& \eta^{\rm ps} \cos(\frac{2\pi}{T}t - \frac{2\pi}{3})  \label{eta2}
\end{eqnarray}
where $\eta^{\rm ps}$ is an undetermined time-independent coefficient corresponding to $c_1$ in Eq.(\ref{xi}).
From Eqs.(\ref{eta1})-(\ref{eta2}), one can show that
\begin{eqnarray}  \label{etasq}
(\eta^{\rm ps})^2 = \frac{4}{3} [ (\eta^{\rm ps}_1(t))^2 + (\eta^{\rm ps}_2(t))^2 + \eta^{\rm ps}_1(t) \eta^{\rm ps}_2(t) ]
\end{eqnarray}
Suppose the system is now near the phase boundary from NEPS to uNESS, i.e. $g \rightarrow -3^{-}$, and
the current $\tilde \eta(t)$ is approaching $\eta^{\rm ps}$ in time. Then
\begin{eqnarray}  \label{etatilde}
\frac{d}{dt}(\tilde \eta(t))^2 &\simeq& \frac{4}{3} \frac{d}{dt} [ (\tilde\eta_1(t))^2 + (\tilde\eta_2(t))^2
+ \tilde\eta_1(t) \tilde\eta_2(t) ] \nonumber \\
&=& \frac{4}{3} [ (2\tilde\eta_1+\tilde\eta_2)\frac{d \tilde\eta_1}{dt}
+ (\tilde\eta_1+2 \tilde\eta_2)\frac{d \tilde\eta_2}{dt} ]   \nonumber \\
\end{eqnarray}
Near the phase boundary, $\tilde\eta_i$ is small. Expand $A_i(\frac{1}{3}+\tilde\eta_1,\frac{1}{3}+\tilde\eta_2)$
from Eq.(\ref{Ai}) up to $O((\tilde\eta_i)^3)$, which gives
\begin{widetext}
\begin{eqnarray}  \label{deta1}
\frac{d\tilde\eta_1}{dt} &=& -\frac{1}{4}(g+2p+2)\tilde\eta_1 -(p-\frac{1}{2})\tilde\eta_2
+ \frac{g}{4}(3p-2)(\tilde\eta_1)^2 + \frac{g}{2}(3p-1)\tilde\eta_1 \tilde\eta_2 + \frac{g}{4}(\tilde\eta_2)^2 \nonumber \\
&& + \frac{g^3}{16}(\tilde\eta_1)^3 + \frac{g^3}{16}(\tilde\eta_1)^2 \tilde\eta_2 + \frac{g^3}{16}\tilde\eta_1(\tilde\eta_2)^2 \\
\frac{d\tilde\eta_2}{dt} &=& -(q-\frac{1}{2})\tilde\eta_1 -\frac{1}{4}(g+2q+2)\tilde\eta_2
+ \frac{g}{4}(\tilde\eta_1)^2  + \frac{g}{2}(3q-1)\tilde\eta_1\tilde\eta_2 + \frac{g}{4}(3q-2)(\tilde\eta_2)^2 \nonumber \\
&& + \frac{g^3}{16}(\tilde\eta_1)^2\tilde\eta_2 + \frac{g^3}{16}\tilde\eta_1(\tilde\eta_2)^2 + \frac{g^3}{16}(\tilde\eta_2)^3
\label{deta2}
\end{eqnarray}
\end{widetext}

Substitute Eqs.(\ref{deta1})-(\ref{deta2}) into Eq.(\ref{etatilde}), and since $\tilde \eta(t)$ is slowly varying compared with
the oscillation of period $T$, the quasi-static approximation \cite{quasistatic} is applied such that
$\tilde \eta_1^m(t) \tilde \eta_2^n(t) \simeq \tilde \eta^{m+n}(t) \langle \cos^m(\frac{2\pi}{T}t) \cos^n(\frac{2\pi}{T}t - \frac{2\pi}{3}) \rangle_t$.
After some algebra, one have
\begin{eqnarray}
\frac{d}{dt} \tilde \eta^2 = \frac{1}{2} (g+3) \tilde\eta^2 + \frac{3}{32}(-g)^3 \tilde \eta^4
\end{eqnarray}
When $g\lesssim -3$, $\frac{d}{dt} \tilde \eta^2 =0$ gives only one stable fixed point
$\tilde \eta^2 = -\frac{16}{3(-g)^3}(g+3)$,  which gives the oscillation amplitude
\begin{eqnarray}  \label{etaas}
\eta^{\rm ps} = \frac{4}{|g|} |\frac{g+3}{3g}|^\frac{1}{2}
\simeq \frac{4}{9} |g+3|^\frac{1}{2}.\label{hopfamp}
\end{eqnarray}
This analytical result is consistent with the numerical values near the phase boundary,
$g \lesssim -3$, as shown by the dashed fitted curve in Fig.\ref{urn3as01_p0d9_g_vs_amp_invT2.eps}(a) and also in  Fig.\ref{urn3as01_u_p_vs_amp_invT.eps}(a).
Since $\eta^{\rm ps}$ drops to zero continuously from NEPS to uNESS,
$\xi^{\rm ps}(t)$ approaches $\xi^{\rm ss}$ and
$K^{\rm ps}_{i\rightarrow i+1}(t)$ will also approach $K^{\rm ss}$.

\begin{figure}[H]
  \begin{center}
    \includegraphics[width=\columnwidth]{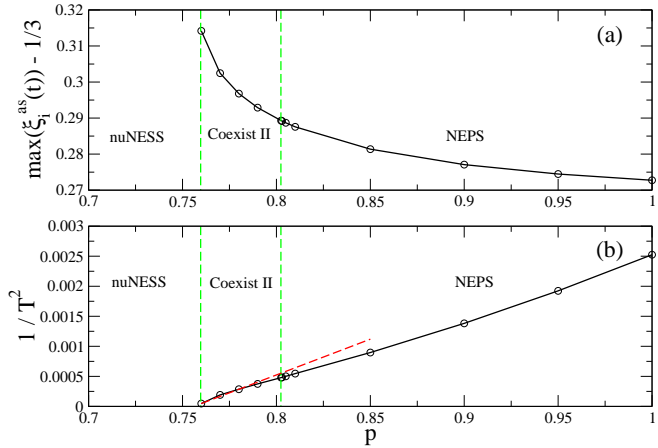}
  \end{center}
  \vspace{-5pt}
  \caption{(a) The amplitude of occupation oscillation, ${\rm max}(\xi^{\rm ps}_i(t))-\frac{1}{3}$,
  is plotted as a function of $p$ at $g=-3.3$ at NEPS.
  (b) The inverse of square of oscillation period, $1/T^2$, is plotted as a function of $p$ at $g=-3.3$ at NEPS.
      Red dashed line is the linear fitting near phase boundary verifying the infinite-period bifurcation characteristics.
   The vertical dashed lines are the phase boundaries. }
  \label{urn3as01_nu_gn3d3_p_vs_amp_invT2.eps}
  \vspace{25pt}
\end{figure}

On the other hand, when the system transits from NEPS to nuNESS,
there appears a coexistence region in between the two phase boundaries as shown in Fig.\ref{phasediagnew.eps}(a).
In particular, the vanishing of the NEPS is via an infinite-period global bifurcation at which the period of oscillation diverges, but the oscillation amplitude remains finite (see Fig.\ref{urn3as01_p0d9_g_vs_amp_invT2.eps}(a))
as the phase boundary is approached, as discussed in previous section. Such a behavior is confirmed in the plot of $1/T^2$ vs. $g$ in Fig.\ref{urn3as01_u_p_vs_amp_invT.eps}(b) for fixed $p=0.9$ (path 2 in Fig.\ref{phasediagnew.eps}(a)), showing a linear behavior (dashed straight line) near the phase boundary.
Fig.\ref{urn3as01_nu_gn3d3_p_vs_amp_invT2.eps} demonstrates a similar behavior characterized by the infinite-period global bifurcation as $p$ is varied with fixed $g=-3.3$ (path 3 in Fig.\ref{phasediagnew.eps}(a)).

\subsection{Non-uniformity and Entropy production}
\begin{figure}[h]
  \begin{center}
    \includegraphics[width=3in]{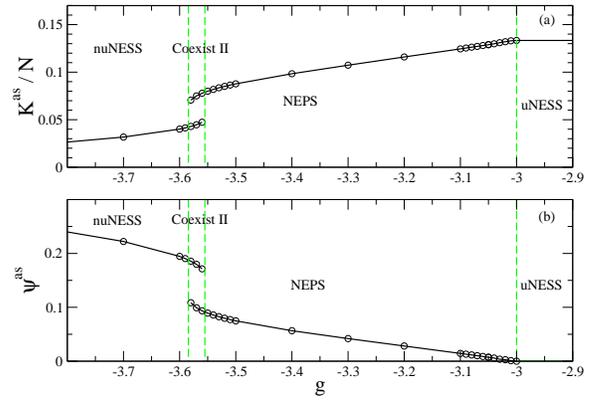}
  \end{center}
  \vspace{-5pt}
  \caption{(a) $K^{\rm as} / N$ are plotted as a function of $g$ at $p$=0.9.
  (b) $\psi^{\rm as} \equiv \langle \psi^{\rm as}(t) \rangle_t$ are plotted as a function of $g$ at $p$=0.9.
  For $g > -3$ at uNESS, $\psi^{\rm as} \equiv \psi^{\rm ss} = 0$.
  The vertical dashed lines are the phase boundaries.}
  \label{urn3as01_p0d9_g_vs_Kavg_psiavg.eps}
  \vspace{25pt}
\end{figure}

\begin{figure}[H]
  \begin{center}
    \includegraphics[width=\columnwidth]{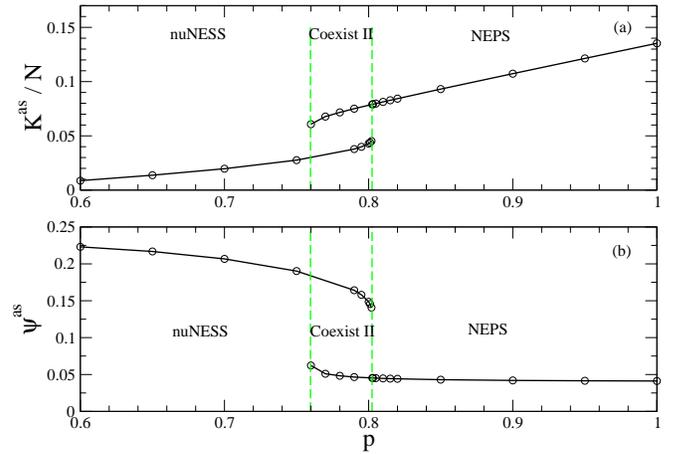}
  \end{center}
  \vspace{-5pt}
  \caption{
  (a) $K^{\rm as} / N$ are plotted as a function of $p$ at $g$=-3.3.
  $K^{\rm as} \equiv \langle K^{\rm as}_{1\rightarrow 2}(t)\rangle_t = \langle K^{\rm as}_{2\rightarrow 3}(t)\rangle_t
  = \langle K^{\rm as}_{3\rightarrow 1}(t)\rangle_t$ and
  (b) $\psi^{\rm as} \equiv \langle \psi^{\rm as}(t) \rangle_t$
  are plotted as a function of $p$ at $g$=-3.3.
  The vertical dashed lines are the phase boundaries.}
  \label{urn3as01_gn3d3_p_vs_Kavg_psiavg.eps}
  \vspace{25pt}
\end{figure}

The mean particle flux  in the NEAS can be measured by  the time average of the flux by $K^{\rm as} \equiv \langle K^{\rm as}_{1\rightarrow 2}(t)\rangle_t = \langle K^{\rm as}_{2\rightarrow 3}(t)\rangle_t
  = \langle K^{\rm as}_{3\rightarrow 1}(t)\rangle_t$. The mean non-uniformity  in the NEAS can also be defined in a similar way
  from Eq.(\ref{nonuniformity}).
Figs.\ref{urn3as01_p0d9_g_vs_Kavg_psiavg.eps} and \ref{urn3as01_gn3d3_p_vs_Kavg_psiavg.eps} show
$K^{\rm as} / N$ and $\psi^{\rm as} \equiv \langle \psi^{\rm as}(t) \rangle_t$ at different $g$ with fixed $p=0.9$
and at different $p$ with fixed $g=-3.3$, which cover NESS, NEPS, and their coexistence.
The mean flux in general increases with $p$ (the driving of the system) and decreases with the inter-particle attraction ($|g|$).
 The mean  non-uniformity in general decreases with $p$  but increases with the inter-particle attraction ($|g|$).

When the system is out of equilibrium, the entropy production does not vanish.
The time average $\langle \frac{dS}{dt}\rangle_t = \langle\frac{dE}{dt}\rangle_t = 0$ for equilibrium state, NESS, and NEPS. According to the thermodynamic law in Eq.(\ref{thermolaw})
and also from Eq.(\ref{dWK}), we arrive at
\begin{eqnarray}  \label{diSKneps}
\left\langle \frac{d_i S}{dt} \right\rangle_t
= 3 K^{\rm as} \log(\frac{p}{q})
\end{eqnarray}
by noticing $\mu=\beta^{-1}\log(\frac{p}{q})$
and $K^{\rm as} \equiv \langle K^{\rm as}_{1\rightarrow 2}(t)\rangle_t = \langle K^{\rm as}_{2\rightarrow 3}(t)\rangle_t
  = \langle K^{\rm as}_{3\rightarrow 1}(t)\rangle_t$.
Here the superscript {\rm as} stands for equilibrium state, NESS, or NEPS.

\begin{figure}[H]
  \begin{center}
    \includegraphics[width=\columnwidth]{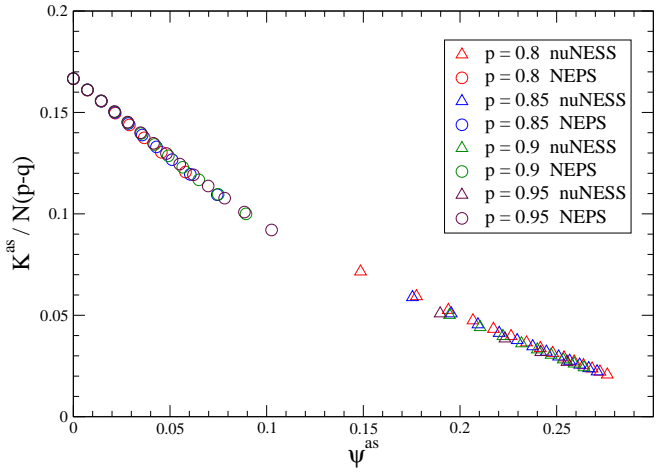}
  \end{center}
  \vspace{-5pt}
  \caption{ The relationship between $\psi^{\rm as} \equiv \langle \psi^{\rm as}(t) \rangle_t$ and
  $\frac{K^{\rm as}}{N(p-q)}$ for different $p$ at different kinds of NEAS.
 The data points fall onto a single curve.  }
  \label{urn3as01_psiavg_vs_Kavg.eps}
  \vspace{25pt}
\end{figure}

Since both the entropy production and non-uniformity depends on the  degree of non-equilibrium or orderliness of the system, one may suspect they are related. As shown in Fig.\ref{urn3as01_psiavg_vs_Kavg.eps}, when the relationship between
$\psi^{\rm as}$ and $\frac{K^{\rm as}}{N(p-q)}$ for different $p$ at different kinds of NEAS are plotted,
all data  collapsed into a single universal curve.
It suggests a general relation
\begin{eqnarray}
\left\langle \frac{d_i S}{dt} \right\rangle_t = N \Phi(\psi^{\rm as}) (p-q)\log(\frac{p}{q})\label{FDR}
\end{eqnarray}
where the function $\Phi(\psi^{\rm as})$ is some decreasing function,
i.e., $\Phi'(\psi^{\rm as}) < 0$.
It is a generalization of the same entropy-production-nonuniformity relation
from NESS \cite{cheng21} to NEPS.

Furthermore, by rewriting Eq.(\ref{FDR}) as
\begin{equation}\left\langle \frac{d_i S}{dt} \right\rangle_t = h\left(\left|p-\dfrac{1}{2}\right|\right)/(1/\Phi),
\quad h(x)\equiv 2Nx\log\left(\frac{\frac{1}{2}+x}{\frac{1}{2}-x}\right),\label{FDR2}
\end{equation}
it contains the flavor of a generalized fluctuation-dissipation relation for NEAS when compared to the usual  fluctuation-dissipation relation relating the damping coefficient and diffusion constant at equilibrium: $\gamma=\frac{k_BT}{D}$. Here the mean entropy production is the dissipation and $1/\Phi$ increases with the non-uniformity and hence is a measure of the fluctuation of the NEAS. The driving source for maintaining the NEAS here is $h(|p-\frac{1}{2}|)$, instead of
 the thermal energy $k_BT$ for the equilibrium case.

%%%%%%%%%%%%%%%%%%%%%%%%%%%%%%%%%%%%%%%%%%%%%%%%%%%%%%%%%%%%%%%%%%%%%%%%%%%%%%%%%%%%%%%%%%%%%%%%%%%
\section{V. Fluctuation Effects}
In the previous sections, only the optimal point solutions are discussed.
However, near the phase boundary, the fluctuation effect may be significant
(c.f. during continuous phase transition at equilibrium).

Let $\vec y \equiv \vec x - \vec \xi(t)$.
From Eq.(\ref{saddle2}),
$\rho(\vec y, t) \propto \exp [ N \sum_{i,j=1}^{M-1} c_{ij}(t) y_i y_j ]$.
Notice that $\sum_{i=1}^M y_i = 0$, $\langle y_i \rangle =0$, and from Eq.(\ref{nonuniformity}), one can decompose the non-uniformity into the mean and fluctuating parts:
\begin{eqnarray}
\psi(t) &=& \frac{1}{M(M-1)} \sum_{(i<j)=1}^M (\xi_i(t) - \xi_j(t))^2
\nonumber \\
&& + \frac{2}{M-1} \sum_{(i\leq j)=1}^{M-1} \langle y_i y_j \rangle
\nonumber \\
&=& \frac{1}{M(M-1)} \sum_{(i<j)=1}^M (\xi_i(t) - \xi_j(t))^2
\nonumber \\
&& + \frac{1}{N} \frac{2}{M-1} \sum_{(i\leq j)=1}^{M-1} ({\bf c}^{-1})_{ij}(t)
\nonumber \\
&\equiv & \psi^{(0)}(t) + \frac{1}{N} \psi^{(1)}(t)
\end{eqnarray}
where $\psi^{(0)}$ is the thermodynamic limit and $\psi^{(1)}$ is the fluctuation.
Take three urns, $M=3$, as an example. When the system approaches the phase boundary from uNESS,
$\vec \xi^{\rm ss} = (\frac{1}{3}, \frac{1}{3}, \frac{1}{3})$, $\psi^{{\rm ss} (0)} = 0$,
$\psi^{{\rm ss} (1)} = -\frac{2}{g+3}$ which diverges at $g\rightarrow -3^{+}$.
However, the non-uniformity itself in thermodynamic limit still converges to zero,
$\lim_{g\rightarrow -3^{+}} \lim_{N\rightarrow \infty} \psi(t) = 0$,
showing that the fluctuation effect doesn't take place in non-uniformity.

To capture the fluctuation effect in observation, we consider the net particle flow fluctuation
$\frac{1}{N}\langle (\delta K_{i \rightarrow i+1}(\vec x))^2 \rangle$.
Notice the continuous form of Eq.(\ref{netflow1}) is
\begin{eqnarray}
K_{i\rightarrow i+1}(t) &=& N \int d^{M-1}y \rho(\vec y,t) \tilde{K}_{i\rightarrow i+1}(\vec y, t)
\end{eqnarray}
where
\begin{eqnarray}
&& \tilde{K}_{i\rightarrow i+1}(\vec y,t) \nonumber \\
&=& \frac{p(y_i+\xi_i) {\rm e}^{g(y_i+\xi_i)} -q(y_{i+1}+\xi_{i+1}){\rm e}^{g(y_{i+1}+\xi_{i+1})} }
{{\rm e}^{g(y_i+\xi_i)}+{\rm e}^{g(y_{i+1}+\xi_{i+1})}}, \nonumber
\end{eqnarray}
then the net flow fluctuation
\begin{eqnarray}  \label{netflowfluctuation}
&& \frac{1}{N}\langle (\delta K_{i \rightarrow i+1}(\vec x))^2 \rangle  \nonumber \\
&=& \frac{1}{N} \left(
\langle K_{i\rightarrow i+1}^2(\vec x) \rangle - \langle K_{i\rightarrow i+1}(\vec x) \rangle^2
\right) \nonumber \\
&=&  N((\partial_i \tilde{K}_{i\rightarrow i+1}(0))^2 \langle y_i^2 \rangle
+ (\partial_{i+1} \tilde{K}_{i\rightarrow i+1}(0))^2 \langle y_{i+1}^2 \rangle  \nonumber \\
&& + 2 (\partial_i \tilde{K}_{i\rightarrow i+1}(0)) (\partial_{i+1}\tilde{K}_{i\rightarrow i+1}(0)) \langle y_i y_{i+1} \rangle ) + O(\frac{1}{N})  \nonumber \\
&=& -(\partial_i \tilde{K}_{i\rightarrow i+1}(0))^2 ({\bf c}^{-1})_{ii}
- (\partial_{i+1} \tilde{K}_{i\rightarrow i+1}(0))^2 ({\bf c}^{-1})_{i+1,i+1} \nonumber \\
&& - 2 (\partial_i \tilde{K}_{i\rightarrow i+1}(0)) (\partial_{i+1}\tilde{K}_{i\rightarrow i+1}(0)) ({\bf c}^{-1})_{i,i+1} + O(\frac{1}{N}),
\nonumber \\
\end{eqnarray}
where
\begin{eqnarray}
\partial_i \tilde{K}_{i\rightarrow i+1}(0) &=& \frac{p}{{\rm e}^{-g(\xi_i-\xi_{i+1})}+1} \nonumber \\
&& + \frac{p g \xi_i}{({\rm e}^{-g(\xi_i-\xi_{i+1})}+1)({\rm e}^{-g(\xi_{i+1}-\xi_i)}+1)} \nonumber \\
\partial_{i+1} \tilde{K}_{i\rightarrow i+1}(0) &=& - \frac{q}{{\rm e}^{-g(\xi_{i+1}-\xi_i)}+1} \nonumber \\
&& - \frac{q g \xi_{i+1}}{({\rm e}^{-g(\xi_i-\xi_{i+1})}+1)({\rm e}^{-g(\xi_{i+1}-\xi_i)}+1)}. \nonumber
\end{eqnarray}

In particular, for $M=3$, when the system approaches to the phase boundary from uNESS, $g > -3$, from Appendix B, one has
\begin{eqnarray} \label{css}
({\bf c}^{\rm ss})^{-1} = -\frac{2}{3(g+3)}\left(
  \begin{array}{cc}
  2 & -1 \\
  -1 & 2 \\
  \end{array}
  \right)
  \end{eqnarray}
and hence
\begin{eqnarray}  \label{netflowfluctss}
\frac{1}{N}\langle (\delta K^{\rm ss}_{i \rightarrow i+1}(\vec x))^2 \rangle
= \frac{(1-pq)}{24} \frac{1}{g+3} + O(\frac{1}{N})
\end{eqnarray}
which is divergent as $g\rightarrow -3^{+}$.

To get the net flow fluctuation at NEPS, one have to first solve for $({\bf c}^{\rm ps})^{-1}$
from Eq.(\ref{fluctutation1}). Near the phase boundary from NEPS, at $g \rightarrow -3^{-}$,
$|\xi^{\rm ps}_i(t)-\frac{1}{3}|=|\eta^{\rm ps}_i(t)|\ll 1$.
Expand $A_i(\xi^{\rm ps}_1(t),\xi^{\rm ps}_2(t))$ around $(\frac{1}{3},\frac{1}{3})$ up to
$O((\eta^{\rm ps})^3)$, and then apply quasi-static approximation, one would get
\begin{widetext}
\begin{eqnarray}
{\bf a}^{\rm ps}
&\equiv&
  \left(
  \begin{array}{cc}
  \partial_1 A_1(\xi^{\rm ps}_1(t),\xi^{\rm ps}_2(t)) & \partial_2 A_1(\xi^{\rm ps}_1(t),\xi^{\rm ps}_2(t)) \\
  \partial_1 A_2(\xi^{\rm ps}_1(t),\xi^{\rm ps}_2(t)) & \partial_2 A_2(\xi^{\rm ps}_1(t),\xi^{\rm ps}_2(t)) \\
  \end{array}
  \right)
  \nonumber \\
&\simeq&
  \left(
  \begin{array}{cc}
  -\frac{1}{4}(g+2p+2)+\frac{3}{32}g^3(\eta^{\rm ps})^2 & -\frac{1}{2}(p-q) \\
  \frac{1}{2}(p-q) & -\frac{1}{4}(g+2q+2)+\frac{3}{32}g^3(\eta^{\rm ps})^2 \\
  \end{array}
  \right)
  \nonumber \\
  &\simeq& -\frac{1}{4}
  \left(
  \begin{array}{cc}
  |g+3| + (p-q) & 2(p-q) \\
  -2(p-q) & |g+3| - (p-q) \\
  \end{array}
  \right)
%&\simeq& -\frac{1}{4}
%\left(
%  \begin{array}{cc}
%  -\epsilon + (p-q) & 2(p-q) \\
%  -2(p-q) & -\epsilon - (p-q) \\
%  \end{array}
%  \right)
\end{eqnarray}
\end{widetext}
and assume
\begin{eqnarray}
{\bf b}^{\rm ps}
&\equiv&
  \left(
  \begin{array}{cc}
  B_{11}(\xi^{\rm ps}_1(t),\xi^{\rm ps}_2(t)) & B_{12}(\xi^{\rm ps}_1(t),\xi^{\rm ps}_2(t)) \\
  B_{21}(\xi^{\rm ps}_1(t),\xi^{\rm ps}_2(t)) & B_{22}(\xi^{\rm ps}_1(t),\xi^{\rm ps}_2(t)) \\
  \end{array}
  \right)
  \nonumber \\
&\simeq&
  \left(
  \begin{array}{cc}
  B_{11}(\frac{1}{3},\frac{1}{3}) & B_{12}(\frac{1}{3},\frac{1}{3}) \\
  B_{21}(\frac{1}{3},\frac{1}{3}) & B_{22}(\frac{1}{3},\frac{1}{3}) \\
  \end{array}
  \right)
\nonumber \\
&=& \frac{1}{6}
  \left(
  \begin{array}{cc}
  2 & -1 \\
  -1 &2 \\
  \end{array}
  \right)
\end{eqnarray}
By quasi-static approximation, $\frac{d}{dt}({\bf c}^{\rm ps})^{-1}=0$, then Eq.(\ref{fluctutation1})
is reduced to ${\bf a}^{\rm ps} ({\bf c}^{\rm ps})^{-1} + ({\bf c}^{\rm ps})^{-1} ({\bf a}^{\rm ps})^{\rm t} = 2 {\bf b}^{\rm ps} $,
which gives
\begin{eqnarray}  \label{cas}
({\bf c}^{\rm ps})^{-1}
= -\frac{2}{3|g+3|}
  \left(
  \begin{array}{cc}
  2 & -1 \\
  -1 &2 \\
  \end{array}
  \right)
\end{eqnarray}
%Notice that it is non-analytic at $g=-3$ \cite{nonanalytic}.
From Eq.(\ref{netflowfluctuation}), one get
\begin{eqnarray} \label{netflowfluctas}
\frac{1}{N}\langle \langle (\delta K^{\rm ps}_{i \rightarrow i+1}(\vec x))^2 \rangle \rangle_t
= \frac{(1-pq)}{24} \frac{1}{|g+3|} + O(\frac{1}{N})
\end{eqnarray}
which is divergent as $g\rightarrow -3^{-}$.
%The divergence at $g=-3$ from Eqs.(\ref{netflowfluctss}) and (\ref{netflowfluctas})
%can be verified by Monte Carlo simulation as shown in Fig.xx

In previous section, at NEPS, one gets the oscillation amplitude of the occupation number,
which is equivalent to $\eta^{\rm ps}$ expressed in Eq.(\ref{etaas}). This amplitude is derived from
dynamical equation in Eq.(\ref{dynamics1}). However, it is also interesting to study its thermal fluctuation effect,
especially near phase transition. Define the stochastic variable
\begin{eqnarray}
\eta^2(t) \equiv \frac{4}{3} [ (x_1-\frac{1}{3})^2 + (x_2-\frac{1}{3})^2 + (x_1-\frac{1}{3})(x_2-\frac{1}{3}) ]  \nonumber \\
\end{eqnarray}
Notice that it is different from the deterministic variable $\tilde \eta(t)$ in Eq.(\ref{etatilde}).
In the following, we are going to show that $\langle \langle \eta^2(t) \rangle \rangle_t = (\eta^{\rm ps})^2$
in the thermodynamic limit.

Let $y_i\equiv x_i - \xi^{\rm ps}_i = x_i - \frac{1}{3} - \eta^{\rm ps}_i$, then
\begin{eqnarray}
&& \langle \langle \eta^2(t) \rangle \rangle_t   \nonumber \\
&=& \frac{4}{3} \langle \langle (\eta^{\rm ps}_1+y_1)^2 + (\eta^{\rm ps}_2+y_2)^2 + (\eta^{\rm ps}_1+y_1)(\eta^{\rm ps}_2+y_2) \rangle \rangle_t
\nonumber \\
&=& \frac{4}{3} \langle (\eta^{\rm ps}_1)^2 + (\eta^{\rm ps}_2)^2 + \eta^{\rm ps}_1\eta^{\rm ps}_2 \rangle_t
+ \frac{4}{3}\langle y_1^2+y_2^2+y_1y_2 \rangle  \nonumber \\
&=& (\eta^{\rm ps})^2 + O(\frac{1}{N}).
\end{eqnarray}
Its fluctuation is given by
\begin{eqnarray}
&& N \langle \langle \delta \eta^2(t) \rangle \rangle_t \nonumber \\
&\equiv& N ( \langle\langle \eta^4(t) \rangle\rangle_t - \langle\langle \eta^2(t) \rangle\rangle_t^2 ) \nonumber \\
&=& \frac{8}{3}(\eta^{\rm ps})^2 \langle y_1^2+y_2^2+y_1y_2 \rangle +O(\frac{1}{N^2})  \nonumber \\
&=& \frac{128}{243} +O(\frac{1}{N^2}).
\end{eqnarray}
Thermal fluctuation would broaden the width of oscillation amplitude, but the width still keep a finite
constant value near the phase boundary at $g\rightarrow -3^-$. For comparison, when the system approaches
the phase boundary from uNESS, $g\rightarrow -3^+$, $N \langle \langle \delta \eta^2(t) \rangle \rangle_t = 0$.
%The result is verified by Monte Carlo simulation as shown in Fig.xx

Finally, the relation between the dynamical and thermodynamic instabilities are illustrated.
At NESS, the dynamical stability condition is that
all real parts of eigenvalues of $\bf a$  are negative.
Or equivalently, the largest real parts of eigenvalues, ${\rm Re}(\lambda_1({\bf a}))$, is negative.
Similarly, the thermodynamic stability condition is that
all eigenvalues of $\bf c$ are negative. Or equivalently,
the largest eigenvalues, $\lambda_1({\bf c})$, is negative.

\begin{figure}[h]
  \begin{center}
    \includegraphics[width=\columnwidth]{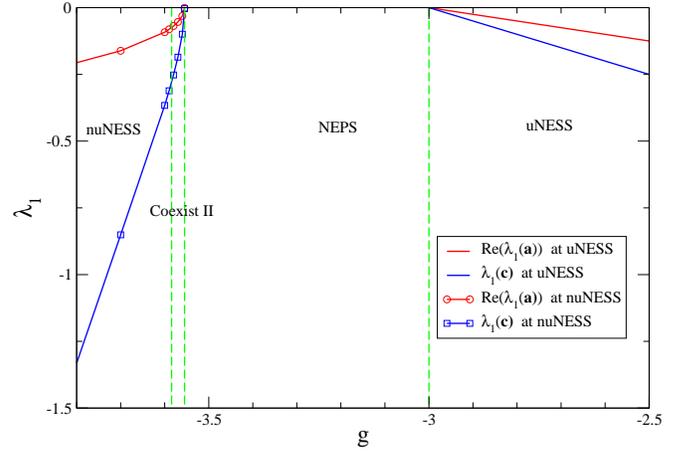}
  \end{center}
  \vspace{-5pt}
  \caption{The largest real parts of eigenvalues of $\bf a$, ${\rm Re}(\lambda_1({\bf a}))$,
    and the largest eigenvalues of $\bf c$, $\lambda_1({\bf c})$,
  are plotted as a function of $g$ at $p=0.9$. The vertical dashed lines are the phase boundaries.
  The graph corresponds to the path 1 and 2 in Fig.\ref{phasediagnew.eps}(a). }
  \label{urn3as01_p0d9_g_vs_lambda.eps}
\end{figure}

Figs.\ref{urn3as01_p0d9_g_vs_lambda.eps} and \ref{urn3as01_nu_gn3d3_p_vs_lambda.eps}
show ${\rm Re}(\lambda_1({\bf a}))$ and $\lambda_1({\bf c})$ at NESS until it meets the phase boundaries.
It is found that both ${\rm Re}(\lambda_1({\bf a}))$ and $\lambda_1({\bf c})$ approach to zero from below.
It implies the dynamical and thermodynamic instabilities occur simultaneously for both uNESS and nuNESS. This result
is consistent with the theorem presented in Appendix B. Thus although we did not prove in general that thermodynamic stability implies dynamic stability, our example for the NESSs in the 3-urn model supports the validity of the above assertion. In other words, the equivalence of dynamic and thermodynamic stability is demonstrated in the 3-urn uNESS and nuNESS.
\begin{figure}[h]
  \begin{center}
    \includegraphics[width=3in]{urn3as01_nu_gn3d3_p_vs_lambda.eps}
  \end{center}
  \vspace{-5pt}
  \caption{The largest real parts of eigenvalues of $\bf a$, ${\rm Re}(\lambda_1({\bf a}))$,
    and the largest eigenvalues of $\bf c$, $\lambda_1({\bf c})$,
  are plotted as a function of $p$ at $g=-3.3$. The vertical dashed line is the phase boundary.
  The graph corresponds to the path 3 in Fig.\ref{phasediagnew.eps}(a). }
  \label{urn3as01_nu_gn3d3_p_vs_lambda.eps}
  \vspace{25pt}
\end{figure}
%%%%%%%%%%%%%%%%%%%%%%%%%%%%%%%%%%%%%%%%%%%%%%%%%%%%%%%%%%%%%%%%%%%%%%%%%%%%%%%%%%%%%%%%%%%%%%%%%%%

\section{VI. conclusion}
The non-equilibrium asymptotic state (NEAS) is proposed and formulated in the framework
of Fokker-Planck equation under WKB approximation (thermodynamic limit).
NEAS is a generalization of non-equilibrium steady state (NESS) which also include all other kinds of asymptotic states,
such as equilibrium state, non-equilibrium periodic state (NEPS), non-equilibrium quasi-periodic state (NEQPS), and non-equilibrium chaotic state (NECS).

In our framework, the dynamics and thermodynamics of the system are not treated independently,
but are mutually connected. It is shown that the dynamical stable NESS is always thermodynamic stable.
Dynamical stability condition is that the real part of all eigenvalues of
a are negative and thermodynamic stability condition is that all eigenvalues of c are negative.
For uniform NESS in $M$ urn ring model, both dynamical and thermodynamic stability criteria are proved to be equivalent. For non-uniform NESS in 3-urn model, the above equivalence is also demonstrated.
The NEPS is constructed in our framework and also illustrated in the
Ehrenfest urn ring model and their time-average properties are calculated explicitly. Its physical properties such as
the dissipation-nonuniformity relation, and its transitions to non-equilibrium steady state (NESS)
are illustrated.

Only the 3-urn model is considered in this paper and the possible NEAS is limited to EQ, NESS, and NEPS. Larger number of urns on a ring will lead to a higher dimensional phase space which allows the possibility of more complex dynamics such as NEQPS and NECS, which is under current studies. Our theoretical framework can be extended to investigate the NECS that may enable one to distinguish chaotic and stochastic fluctuation in complex thermodynamic nonlinear system.

\section{acknowledgments}
 This work  has been supported by the National Science and Technology Council of Taiwan under grants nos. 110-2112-M-008-026-MY3 (PYL) and 111-2112-M-018-005 (CHC).

\appendix*

\section{Appendix A: Asymptotic State Solution of Multivariate Linear Fokker-Planck Equation}
The multivariate linear Fokker-Planck Equation of dimension $D$ for asymptotic state reads
\begin{widetext}
\begin{eqnarray} \label{fpe3}
   \frac{\partial \rho^{\rm as}(\vec x,t)}{\partial t}
  = -\sum_{i=1}^D \frac{\partial}{\partial x_i}\left[ \left( A_i(\vec\xi(t),t)+\sum_{j=1}^D a_{ij}(t)(x_j-\xi_j(t))\right)
  \rho^{\rm as}(\vec x,t)\right]
    + \frac{1}{2 N}\sum_{i,j=1}^D b_{ij}(t)  \frac{\partial^2 }{\partial x_i \partial x_j}
\rho^{\rm as}(\vec x,t)
\end{eqnarray}
\end{widetext}
where $\bf a$ and $\bf b$ are matrices of dimension $D\times D$. $\bf b$ is symmetric.
In large $N$ limit, the form of the solution becomes
\begin{widetext}
\begin{eqnarray}
\rho^{\rm as}(\vec x, t) = \left(\frac{N}{\pi}\right)^\frac{D}{2} \det{(-{\bf c})^\frac{1}{2}}
\exp\left[ N \sum_{i,j=1}^D c_{ij}(t) (x_i-\xi_i(t))(x_j-\xi_j(t)) \right]
\end{eqnarray}
\end{widetext}
where $\bf c$ is a symmetric matrix determined by $\bf a$ and $\bf b$.
Substitute this form into Eq.(\ref{fpe3}) and keep the leading order in $N$, we get
\begin{eqnarray}
(\partial_t {\vec \xi})^{\rm t} {\bf c} {\vec y} &=& {\vec A}^{\rm t} {\bf c} {\vec y} \label{dxidt} \\
{\vec y}^{\rm t} (\partial_t {\bf c}) {\vec y} &=& -2 {\vec y}^{\rm t} {\bf c a} {\vec y}
+ 2 {\vec y}^{\rm t} {\bf c b c} {\vec y} \label{dcdt}
\end{eqnarray}
where ${\vec y} \equiv {\vec x}-{\vec \xi(t)}$ for any $\vec x$.
Eq.(\ref{dxidt}) is reduced to
\begin{eqnarray}
\partial_t {\vec \xi} &=& {\vec A}
\end{eqnarray}
Notice that $\bf cbc$ is symmetric but $\bf ca$
is not necessary to be. Using ${\vec y}^{\rm t} {\bf c a} {\vec y} = {\vec y}^{\rm t} {\bf a^{\rm t} c } {\vec y}$,
Eq.(\ref{dcdt}) can be rewritten as
\begin{eqnarray}
{\vec y}^{\rm t} (\partial_t {\bf c}) {\vec y} = - {\vec y}^{\rm t} ({\bf c a} + {\bf a^{\rm t}}c ) {\vec y}
+ 2 {\vec y}^{\rm t} {\bf c b c} {\vec y}
\end{eqnarray}
which gives
\begin{eqnarray}
\partial_t {\bf c}  = - {\bf c a} - {\bf a^{\rm t}}c  + 2 {\bf c b c}
\end{eqnarray}
or equivalently,
\begin{eqnarray}
\partial_t {\bf c}^{\rm -1}  = {\bf a} {\bf c}^{-1} + {\bf c}^{-1} {\bf a}^{\rm t} - 2 {\bf b}
\end{eqnarray}
which is a system of first-order differential equations of dimension $D(D+1)/2$ to uniquely
determine the same number of independent matrix elements of $\bf c$.

\section{Appendix B: Theorem for the Dynamical and Thermodynamic Stability of NESS}
{\it Lemma} (Lyapunov Theorem \cite{antsaklis}): Given the system of linear equation (Lyapunov equation),
${\bf a} {\bf c}^{-1} + {\bf c}^{-1} {\bf a}^{\rm t} = 2 {\bf b}$, where
$\bf b$ is non-singluar, symmetric, and positive definite,
and all the real parts of eigenvalues of $\bf a$ is negative, ${\rm Re}(\lambda({\bf a})) < 0$,
then all the eigenvalues of $\bf c$ are real and negative, $\lambda({\bf c}) < 0$.

 {\it Proof}: Notice that ${\bf c}^{-1}$ can be expressed  as
 \begin{eqnarray}
 {\bf c}^{-1} = -2 \int_0^\infty d\tau {\rm e}^{\tau \bf a} {\bf b} {\rm e}^{\tau {\bf a}^{\rm t}}
 \end{eqnarray}
 which can checked by
 \begin{eqnarray}
 {\bf a} {\bf c}^{-1} + {\bf c}^{-1} {\bf a}^{\rm t}
 &=& -2 \int_0^\infty d\tau ( {\bf a} {\rm e}^{\tau \bf a} {\bf b} {\rm e}^{\tau {\bf a}^{\rm t}}
 + {\rm e}^{\tau \bf a} {\bf b} {\rm e}^{\tau {\bf a}^{\rm t}} {\bf a}^{\rm t} )  \nonumber \\
 &=& -2  \int_0^\infty d\tau \frac{d}{d\tau} ( {\rm e}^{\tau \bf a} {\bf b} {\rm e}^{\tau {\bf a}^{\rm t}} )  \nonumber \\
 &=& -2 \left[ {\rm e}^{\tau \bf a} {\bf b} {\rm e}^{\tau {\bf a}^{\rm t}} \right]^\infty_0  \nonumber \\
 &=& 2 {\bf b}
 \end{eqnarray}
if the condition ${\rm Re}(\lambda({\bf a})) < 0$ is imposed. Since $\bf b$ is non-singular, then
\begin{eqnarray} \label{lyapunov2}
{\bf c} = -2 \int_0^\infty d\tau {\rm e}^{-\tau{\bf a}^{\rm t}} {\bf b}^{-1} {\rm e}^{-\tau \bf a}
\end{eqnarray}
Using the fact that $\bf b$ is symmetric and positive definite (so is ${\bf b}^{-1}$),
we get that $\bf c$ is symmetric and negative definite, which implies
all eigenvalues of $\bf c$ are real and negative. Q.E.D.

At NESS, $\bf a$ and $\bf c$ defined in Eq.(\ref{fpe2}) are connected by the Lyapunov equation
(Eq.(\ref{fluctutation1}) with $\partial_t {\bf c}^{-1}=0$ ).
Dynamical stability condition is that the real part of all eigenvalues of $\bf a$ are negative
and thermodynamic stability condition is that all eigenvalues of $\bf c$ are negative.
By the above lemma,  we have the following theorem.

{\it Theorem}: At NESS, the dynamical stability implies the thermodynamic stability.

In other words, thermodynamic instabililty at NESS leads to dynamical instability.

\section{Appendix C: Stability of Uniform Non-equilibrium Steady State}

The Fokker-Planck Equation in Eq.(\ref{fpe1}) is reduced to Eq.(\ref{fpe2}) in large $N$ limit.
For uniform NESS in $M$ urn ring model, when $M=3$, from Eqs.(23) and (25) in Ref.\cite{cheng21}, we have
\begin{eqnarray}
{\bf a}=-\frac{1}{2}
\left(
  \begin{array}{cc}
    1+p+\frac{g}{2} & p-q \\
    q-p & 1+q+\frac{g}{2} \\
  \end{array}
\right)
\end{eqnarray}
in which its eigenvalues are $-\frac{g+3}{4} \pm i \frac{\sqrt{3}}{4} (p-q)$.

\begin{eqnarray}
{\bf c}=-\frac{g+3}{2}
\left(
  \begin{array}{cc}
    2 & 1 \\
    1 & 2 \\
  \end{array}
\right)
\end{eqnarray}
in which its eigenvalues are $-\frac{g+3}{2}$ and $-\frac{3(g+3)}{2}$.
It can be seen that the real part of eigenvalues of $\bf a$ and the eigenvalues of $\bf c$ are
are negative (positive) if $g>-3$ ($g<-3$).
It implies uNESS in three urn ring model is both dynamical and thermodynamic stable (unstable) if $g>-3$ ($g<-3$).

When $M \geq 4$,
\begin{widetext}
\begin{eqnarray} \label{ness_a}
{\bf a} = -\frac{1}{2}
\left(
  \begin{array}{cccccccc}
    1+p+\frac{3g}{2M} & p-q & p+\frac{g}{2M} & p+\frac{g}{2M} & \cdots
    & p+\frac{g}{2M} & p+\frac{g}{2M} & p+\frac{g}{2M} \\
    -p-\frac{g}{2M} & 1+\frac{g}{M} & -q-\frac{g}{2M} & 0 & \cdots & 0 & 0 & 0 \\
    0 & -p-\frac{g}{2M} & 1+\frac{g}{M} & -q-\frac{g}{2M} & \cdots & 0 & 0 & 0 \\
    0 & 0 & -p-\frac{g}{2M} & 1+\frac{g}{M} & \cdots & 0 & 0 & 0 \\
    \vdots & \vdots & \vdots & \vdots & \ddots & \vdots & \vdots & \vdots \\
    0 & 0 & 0 & 0 & \cdots & 1+\frac{g}{M} & -q-\frac{g}{2M} & 0 \\
    0 & 0 & 0 & 0 & \cdots & -p-\frac{g}{2M} & 1+\frac{g}{M} & -q-\frac{g}{2M} \\
    q+\frac{g}{2M} & q+\frac{g}{2M} & q+\frac{g}{2M} & q+\frac{g}{2M} & \cdots
    & q+\frac{g}{2M} & q-p & 1+q+\frac{3g}{2M} \\
  \end{array}
\right)
\end{eqnarray}
\end{widetext}
and
\begin{eqnarray} \label{ness_b}
{\bf b} = \frac{1}{2M}
\left(
  \begin{array}{ccccccc}
    2 & -1 & 0 & \cdots & 0 & 0 & 0 \\
    -1 & 2 & -1 & \cdots & 0 & 0 & 0 \\
    0 & -1 & 2 & \cdots & 0 & 0 & 0 \\
    \vdots & \vdots & \vdots & \ddots & \vdots & \vdots & \vdots \\
    0 & 0 & 0 & \cdots & 2 & -1 & 0 \\
    0 & 0 & 0 & \cdots & -1 & 2 & -1 \\
    0 & 0 & 0 & \cdots & 0 & -1 & 2 \\
  \end{array}
\right)
\end{eqnarray}

The probability density in Eq.(\ref{saddle2}) is known if
the matrix $\bf c$ is solved from the Lyapunov equation
$ {\bf a} {\bf c}^{-1} + {\bf c}^{-1} {\bf a}^{\rm t} = 2 {\bf b}$, which gives
\begin{eqnarray}
{\bf c}=-\frac{2(g+M)}{M}
\left(
  \begin{array}{cccc}
    2 & 1 & \cdots & 1 \\
    1 & 2 & \cdots & 1 \\
    \vdots & \vdots & \ddots & \vdots \\
    1 & 1 & \cdots & 2 \\
  \end{array}
\right)
\end{eqnarray}
for $M\geq 4$. The eigenvalues of the matrix
$\left( \begin{array}{cc} . & . \\ . & . \\  \end{array} \right)$
are $M$ and 1 ($M-2$ degeneracy). Then all eigenvalues of $\bf c$ are negative (positive) if $g>-M$ ($g<-M$).
It implies that uniform NESS is thermodynamic stable (unstable) if $g>-M$ ($g<-M$).

To check the dynamical stability of the uniform NESS, we decompose $\bf a$ in Eq.(\ref{ness_a}) into
\begin{widetext}
\begin{eqnarray}
{\bf a} &=& -\frac{g+M}{4M}
\left(
  \begin{array}{cccccccc}
    3 & 0 & 1 & 1 & \cdots & 1 & 1 & 1 \\
    -1 & 2 & -1 & 0 & \cdots & 0 & 0 & 0 \\
    0 & -1 & 2 & -1 & \cdots & 0 & 0 & 0 \\
    0 & 0 & -1 & 2 & \cdots & 0 & 0 & 0 \\
    \vdots & \vdots & \vdots & \vdots & \ddots & \vdots & \vdots & \vdots \\
    0 & 0 & 0 & 0 & \cdots & 2 & -1 & 0 \\
    0 & 0 & 0 & 0 & \cdots & -1 & 2 & -1 \\
    1 & 1 & 1 & 1 & \cdots
    & 1 & 0 & 3 \\
  \end{array}
\right)
-\frac{p-q}{4}
\left(
  \begin{array}{cccccccc}
    1 & 2 & 1 & 1 & \cdots & 1 & 1 & 1 \\
    -1 & 0 & 1 & 0 & \cdots & 0 & 0 & 0 \\
    0 & -1 & 0 & 1 & \cdots & 0 & 0 & 0 \\
    0 & 0 & -1 & 0 & \cdots & 0 & 0 & 0 \\
    \vdots & \vdots & \vdots & \vdots & \ddots & \vdots & \vdots & \vdots \\
    0 & 0 & 0 & 0 & \cdots & 0 & 1 & 0 \\
    0 & 0 & 0 & 0 & \cdots & -1 & 0 & 1 \\
    -1 & -1 & -1 & -1 & \cdots & -1 & -2 & -1 \\
  \end{array}
\right)
\nonumber \\
&\equiv& -\frac{g+M}{4M} {\bf a}_g  -\frac{p-q}{4} {\bf a}_p
\end{eqnarray}
\end{widetext}
By straightforward computation, it can be seen that all eigenvalues of ${\bf a}_g$
are real and positive, and that of ${\bf a}_p$ are purely imaginary. Notice that
${\bf a}_g$ and ${\bf a}_p$ commute (${\bf a}_g {\bf a}_p = {\bf a}_p {\bf a}_g$), implying that they share the
same set of eigenvectors. Hence,
the real part of eigenvalues of $\bf a$ is equal to the eigenvalues of $-\frac{g+M}{4M} {\bf a}_g$.
Then the real part of all eigenvalues of $\bf a$ are negative (positive) if $g>-M$ ($g<-M$).
It implies that the uniform NESS is dynamical stable (unstable) if $g>-M$ ($g<-M$).
Both the dynamical and thermodynamic stability criteria are equivalent for uniform NESS.
This consequence is consistent with the theorem in Appendix B.

\section{Appendix D: Thermodynamic Law}
We are going to identify the thermodynamic law,
\begin{eqnarray} \label{thermolaw2}
\frac{dS}{dt} = \frac{d_iS}{dt} + \beta \frac{dE}{dt} + \beta \frac{dW}{dt}
\end{eqnarray}
in the Ehrenfest urn ring model of arbitrary $M$.

The Boltzmann entropy of the system is given by
\begin{eqnarray}
S = -\sum_{\vec n} \rho(\vec n,t)
\log \left( \rho(\vec n,t)/\frac{N!}{\prod_{i=1}^M n_i!}
     \right)
\end{eqnarray}
where the multiplication factor $\frac{N!}{\prod_{i=1}^M n_i!}$ is due to the degeneracy of $\rho(\vec n,t)$.
Applying Eq.(\ref{master1}),
the entropy production rate becomes
\begin{widetext}
\begin{eqnarray}
\frac{dS}{dt} &=& -\sum_{\vec n, \vec m}
(W_{\vec n, \vec m} \rho(\vec m,t) - W_{\vec m, \vec n} \rho(\vec n,t))
\log\left(\rho(\vec n,t)/\frac{N!}{\prod_{i=1}^M n_i!}\right) \nonumber \\
&=& \frac{N}{2} \sum_{\vec n, \vec m}
(W_{\vec n, \vec m} \rho(\vec m,t) - W_{\vec m, \vec n} \rho(\vec n,t))
\log\left( \frac{W_{\vec n, \vec m}\rho(\vec m,t)}{W_{\vec m, \vec n}\rho(\vec n,t)} \right)  \nonumber \\
&& +\frac{N}{2} \sum_{\vec n, \vec m}
(W_{\vec n, \vec m} \rho(\vec m,t) - W_{\vec m, \vec n} \rho(\vec n,t))
\log\left( \frac{W_{\vec m,\vec n}}{W_{\vec n,\vec m}}
\frac{\frac{N!}{\prod_{i=1}^M n_i!}}{\frac{N!}{\prod_{i=1}^M m_i!}} \right) \nonumber \\
&=& \frac{d_{\rm i} S}{dt} + \frac{d_{\rm e} S}{dt}
\end{eqnarray}
\end{widetext}
where the first term, $\frac{d_iS}{dt}$, is the internal entropy production rate \cite{schnakenberg},
also known as KL divergence \cite{kldiv}.
By noticing that
\begin{eqnarray}
\frac{W_{\vec m, \vec n}}{W_{\vec n, \vec m}} = \frac{p}{q} \frac{n_i}{n_j+1}
{\rm e}^{\frac{g}{N}(n_i-n_j-1)}
\end{eqnarray}
if the jump is in anti-clockwise (ac) direction, and in clockwise (c) direction,
\begin{eqnarray}
\frac{W_{\vec m, \vec n}}{W_{\vec n, \vec m}} = \frac{q}{p} \frac{n_i}{n_j+1}
{\rm e}^{\frac{g}{N}(n_i-n_j-1)}
\end{eqnarray}
The second term, $\frac{d_eS}{dt}$, can be further re-written as
%\begin{widetext}
\begin{eqnarray}
\frac{d_{\rm e} S}{dt}
&=& \frac{N}{2}
\sum_{\vec n, \vec m} (W_{\vec n, \vec m} \rho(\vec m,t) - W_{\vec m, \vec n} \rho(\vec n,t) )
\frac{g}{N}(n_i-n_j-1)
\nonumber \\
&& + \frac{N}{2}
\sum_{\vec n}{\sum_{\vec m}}^{\rm ac} (W_{\vec n, \vec m} \rho(\vec m,t) - W_{\vec m, \vec n} \rho(\vec n,t))
\log(\frac{p}{q}) \nonumber \\
&& + \frac{N}{2}
\sum_{\vec n}{\sum_{\vec m}}^{\rm c} (W_{\vec n, \vec m} \rho(\vec m,t) - W_{\vec m, \vec n} \rho(\vec n,t))
\log(\frac{q}{p})
\nonumber \\
&=& \sum_{\vec n, \vec m} g [ n_j-(n_i-1)]
W_{\vec m, \vec n} \rho(\vec n,t)  \nonumber \\
&& - N \log(\frac{p}{q}) \sum_{\vec n} {\sum_{\vec m}}^{\rm ac}
(W_{\vec m, \vec n}\rho(\vec n,t)-W_{\vec n, \vec m}\rho(\vec m,t))  \nonumber \\
&=& \beta \frac{dE}{dt} + \beta \frac{dW}{dt}
\end{eqnarray}
%\end{widetext}
The first term is the rate of change of energy
and the second term is the rate of work done by the system
which can also be written as
\begin{eqnarray}
\beta \frac{dW}{dt} &=& -\beta\mu \sum_{i=0}^{M-1} K_{i\rightarrow i+1}
\end{eqnarray}
where $\mu\equiv \beta^{-1} \log (\frac{p}{q})$ is the effective chemical potential difference to actively drive the particle
from one urn to another.
$K_{i\rightarrow i+1}$ is the net particle flow from the $i$-th to the $(i+1)$-th urn, which is defined as
\begin{eqnarray}
K_{i\rightarrow i+1} &\equiv& N \sum_{\vec n}
( W_{(n_i-1,n_{i+1}+1),(n_i,n_{i+1})} \nonumber \\
&& - W_{(n_i+1,n_{i+1}-1),(n_i,n_{i+1})} ) \rho(\vec n,t)
\end{eqnarray}
Now the thermodynamic law in Eq.(\ref{thermolaw2}) is identified.
Note that it holds for general thermodynamic (asymptotic and non-asymptotic states) process.


\begin{references}
\bibitem{saddlepoint}
In equilibrium statistical mechanics, when $N$ is large, this optimal point is usually called
the saddle point. The partition function and any average physical variables
are extended from real to complex plane such that the optimal point in this complex plane
is not locally maximum or minimum, but a saddle point.
This saddle point in the complex plane corresponds to the fixed point
in the parameter space of the Fokker-Planck Equation. To avoid ambiguity to the readers,
we avoid the terminology "saddle point" in the text.
\bibitem{quasistatic}
Quasi-static approximation in the text refers to the case that
the dynamics of the slow variables ($\tilde \eta(t)$ near the phase boundary)
can be approximated by taking the time average of the fast oscillating terms in the equations.
\end{references}
\end{document}